\documentclass[preprint,12pt,number,sort&compress]{elsarticle}

\usepackage[utf8]{inputenc}
\usepackage{amsmath}
\usepackage{graphicx}
\usepackage{xcolor}
\usepackage{soul}
\usepackage{booktabs}
\usepackage{enumitem}
\usepackage{float}
\usepackage{url}
\usepackage{caption}
\usepackage{lineno}

\journal{arXiv}

\begin{document}

\begin{frontmatter}

\title{Molecular details and free energy barriers of ion de-coordination at elevated salinity and pressure and their consequences for membrane separations}

\author[1]{Nathanael S. Schwindt}
\author[2]{Razi Epsztein}
\author[3]{Anthony P. Straub}
\author[4]{Shuwen Yue}
\author[1]{Michael R. Shirts}
\affiliation[1]{organization={Department of Chemical \& Biological Engineering, University of Colorado Boulder},
                city={Boulder},
                state={CO},
                postcode={80309},
                country={USA}}
\affiliation[2]{organization={Faculty of Civil and Environmental Engineering, Technion - Israel Institute of Technology},
                city={Haifa},
                postcode={32000},
                country={Israel}}
\affiliation[3]{organization={Department of Civil, Environmental, and Architectural Engineering, University of Colorado Boulder},
                city={Boulder},
                state={CO},
                postcode={80309},
                country={USA}}
\affiliation[4]{organization={R. F. Smith School of Chemical and Biomolecular Engineering, Cornell University},
                city={Ithaca},
                state={NY},
                postcode={14853},
                country={USA}}
\ead{michael.shirts@colorado.edu}
\cortext[cor1]{Corresponding author.}

\begin{abstract} 

Ion dehydration has been hypothesized to strongly influence separation performance in membrane systems and ion transport in nanoscale channels. However, the molecular details of ion dehydration in membranes are not well understood, in particular under the high pressures and concentrations required for brine treatment.

In this study, we define \textit{de-coordination} as the process by which an ion decreases its total coordination number, including both water molecules and counterions. We estimate the de-coordination free energies in bulk solution for a range of different ions at high pressure and salinity relevant to brine treatment using molecular simulation. We also propose alternatives to the coordination number as the size constraint for traversing nanoscale constrictions, such as the maximum cross-sectional area of the complexed ion.

We show that high operating pressures do not significantly change cation hydration shell stability nor the shell size, while high ionic concentrations lower the free energy barrier to reduce the cation coordination number. We find that anion de-coordination free energies are largely unaffected by elevated salinity and pressure conditions. Finally, we discuss the implications on ion-ion selectivity in separations membranes (e.g. extracting lithium from salt-lake brines) due to the effects of elevated pressure and salinity on ion de-coordination.

\end{abstract}

\begin{graphicalabstract}
\includegraphics[width=\textwidth]{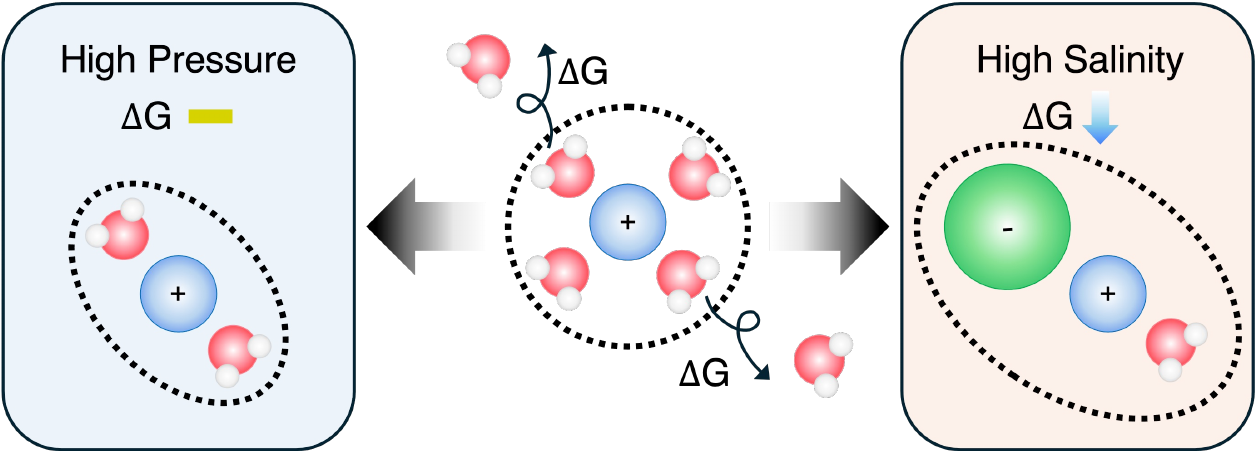}
\end{graphicalabstract}

\begin{highlights}
    \item High salinity decreases cation de-coordination free energies
    \item Elevated pressure does not significantly change de-coordination free energies
    \item We argue that cross-sectional area of the shell is a better metric of hydrated size
    \item Only in a few cases do barriers allow significant discrimination between ions
    \item Stripping ions of more than 2--3 waters is unlikely in polymer membranes
\end{highlights}

\begin{keyword}
Ion dehydration \sep coordination number \sep free energy barriers \sep molecular dynamics \sep biased sampling analysis \sep high pressure \sep high salinity
\end{keyword}

\end{frontmatter}


\section{Introduction}

Efficient and cost-effective brine management is crucial to enable inland desalination and prevent the harmful discharge of saline industrial wastewater.~\cite{davenport_high-pressure_2018,werber_materials_2016,zhou_intrapore_2020}. These brine solutions are common waste streams in seawater desalination, oil and gas production, and industrial manufacturing~\cite{davenport_high-pressure_2018}. Recently, reverse osmosis (RO) and nanofiltration (NF) membranes have shown promise as energy-efficient brine treatment processes when operated at high pressure~\cite{wang_minimal_2020,davenport_high-pressure_2018}. However, a better understanding of ion transport within membranes is necessary to fully unlock the potential of membrane-based brine treatment~\cite{zhou_intrapore_2020,epsztein_towards_2020,faucher_critical_2019}. Brine streams have a high concentration of salts, which requires higher applied pressures to overcome increased osmotic pressures, typically $>$100 bar~\cite{davenport_high-pressure_2018}. Operating at these high pressures has been shown to decrease water permeability and increase salt permeability, which decreases the water-salt selectivity~\cite{davenport_high-pressure_2018,pataroque_salt_2024}. At high salinity, salt permeability increases disproportionally, further decreasing the water-salt selectivity. It has been suggested that easier ion dehydration and reduced electrostatic interactions contribute to this decrease in salt permeability~\cite{chen_transport_2020}. Overall trends in ion rejection have shown the importance of many mechanisms in determining transport, such as steric exclusion, Donnan exclusion, and dielectric exclusion~\cite{saavedra_comparative_2024,freger_dielectric_2023,sun_exploring_2024}. However, these mechanisms do not fully explain trends in ions of similar size and valency.  

\begin{figure*}[ht!]
    \centering
    \includegraphics[width=\textwidth]{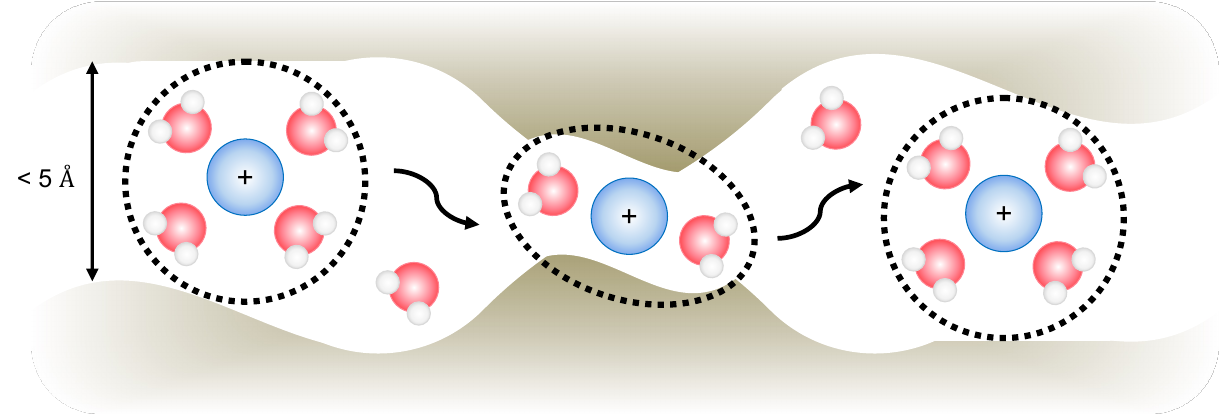}
    \caption{An ion must strip off its tightly coordinated waters and rearrange its coordination shell to traverse nanoscale constrictions in membranes. The volume of the shell must be reduced to move through constrictions. The hydration shell for the blue ion is shown as a dotted black line. }
    \label{fig:dehydration_visualization}
\end{figure*}

Recent work has proposed that ion dehydration can help describe observed trends in ion rejection~\cite{zhou_intrapore_2020,lu_dehydration-enhanced_2023, pavluchkov_indications_2022,chen_steric_2023,zhang_role_2023,meng_enhancing_2024}. These claims largely stem from observed correlations between experimental hydration free energies and trends in ion rejection and energy barriers to permeation, and they have been supported by molecular simulation studies reporting decreased coordination numbers as ions move through nanoscale voids, such as those in RO and NF membranes. Studies of materials such as nanoporous graphene~\cite{sahu_dehydration_2017,sahu_dehydration_2023}, carbon nanotubes (CNTs)~\cite{richards_importance_2012,song_intrinsic_2009}, metal-organic frameworks (MOFs)~\cite{chen_steric_2023,xie_unveiling_2024,xu_perfect_2024}, and graphite sheets~\cite{malani_hydration_2010,malani_effect_2006} have all related changes in ion coordination number with transport properties of these ions. Additionally, recent work using \textit{in situ} time-of-flight secondary ion mass spectrometry (ToF-SIMS) has shown that ions experience a decrease in coordination number as they move through polymer membranes~\cite{chen_steric_2023,lu_situ_2021,lu_dehydration-enhanced_2023}. Figure~\ref{fig:dehydration_visualization} conceptualizes how ion dehydration contributes to ion transport in membrane nanoscale constrictions. The ion must change its coordination shell size to traverse the membrane. 

Although ion dehydration has been implicated in membrane performance for different ions~\cite{chen_steric_2023,epsztein_activation_2019,lu_dehydration-enhanced_2023}, the molecular details of the process are not well understood. Without a clear understanding of how ion dehydration is affected by ion properties and operating conditions, the effects of these variables on the transport cannot be fully understood. Brine treatment membranes are operated at high pressure and salinity, and these conditions have been shown to influence membrane performance~\cite{lim_demystifying_2024,davenport_high-pressure_2018}. We therefore performed a thorough investigation of ion dehydration free energies under relevant conditions for membrane transport and for a range of different ions using molecular simulations to elucidate the molecular details of the ion dehydration process. 

Molecular simulations provide a route to directly study the nanoscale details of ion dehydration. Previous studies of zeolites, graphene, and carbon nanotubes have shown the importance of dehydration in ion transport through nanochannels~\cite{sahu_dehydration_2017,richards_importance_2012,chen_steric_2023,xie_unveiling_2024}. Additional work has described the change in hydration shell structure under nanoscale confinement, similar to the environments within RO or NF membranes~\cite{malani_hydration_2010,malani_effect_2006}. Previous studies have used enhanced sampling techniques to detail the free energy landscape of ion coordination and the free energy barriers to stripping a water molecule from ion coordination~\cite{roy_reaction_2016,brancato_free_2011,song_intrinsic_2009,annapureddy_understanding_2014}. However, to our knowledge, a comprehensive study of ion dehydration at  operating conditions relevant for membrane transport, especially at higher pressures and higher concentrations, does not exist in the literature. 

If ion dehydration is a governing mechanism for ion transport through these membranes, an ion must decrease its complexed size and overcome the associated free energy barrier. Our analysis has implications for a range of nanoscale constrictions, including rigid constrictions like MOFs or CNTs and fluctuating polymer voids or pores. Our discussion focuses on polymer membranes that are common in RO and NF. The prevailing modeling framework for RO and NF membranes is solution-diffusion, while the pore flow framework is often suggested to describe transport in loose NF membranes with slightly larger pores~\cite{wang_pore_2021}. 

Under solution-diffusion, the solute dissolves into the polymer phase and then diffuses through the membrane~\cite{wijmans_solution-diffusion_1995}. This diffusive motion is typically modeled with a single diffusion coefficient that describes the transport of the solute in the membrane. The diffusion coefficient thus accounts for all the free energy barriers a solute must overcome to traverse the polymer membrane. Here, we assume that ion dehydration can be considered one such mechanism included in the diffusion coefficient for transport of aqueous ions.

Under pore flow, the solute moves through pores or channels driven by a pressure gradient, and solutes are excluded by interactions (e.g. electrostatic) with the pore mouth~\cite{wijmans_solution-diffusion_1995,wang_pore_2021}. We assume that ion dehydration is necessary for ions to move through small pores and bottlenecks in the membrane. Therefore, a thorough understanding of ion dehydration is highly relevant for both pictures of transport in RO and NF membranes.

Within a membrane, the physical mechanism of an ion moving through a confined void or constriction involves stripping water molecules from the coordination shell, replacing the stripped water molecules with membrane groups or counterions, or rearranging the coordination shell to allow for additional ion movement, as illustrated in Figure~\ref{fig:dehydration_visualization}. As a result, this process is composed of many free energy barriers. Generally, it is implicitly assumed that the free energy to strip waters from the hydration shell is closely related to the total free energy for this process~\cite{epsztein_activation_2019,richards_experimental_2013,pavluchkov_indications_2022}. The rationale for this is that compensating interactions involving less polar membrane groups would be similar to, but weaker than, the interactions with more polar bulk water. If the strength of all of the ion-ion and ion-membrane interactions are proportional to ion hydration itself, then to a rough approximation, the total free energy barrier to move along the membrane would be proportional to the ion dehydration free energy. In other words, ions with large free energies of dehydration will have concomitantly large barriers to moving through a size-restricted membrane. Under this approximation, we expect that we can gain some insight into the barriers to ion transport in a membrane as a function of species, pressure, and salinity by studying the free energy of dehydration in bulk solution.

The most common definition of dehydration is the process by which an ion is stripped of its tightly coordinated water molecules, or its first hydration shell~\cite{richards_experimental_2013,sahu_dehydration_2017,freger_dielectric_2023}. However, this definition does not account for other species in solution that may surround the ion, such as counterions, which can become significant at lower water coordination, even at moderate concentrations. The total number of molecules moving with the ion, regardless of species, must be low enough that an ion can move through molecularly narrow channels or bottlenecks in the membrane. If there is ion pairing among transporting ions, a change in the number of tightly coordinated water molecules may not significantly change the size of the coordination shell. Therefore, we define de-coordination, rather than dehydration, as the mechanism by which an ion changes its \textit{total} coordination number, and we calculate the de-coordination free energy as the free energy difference between discrete coordination numbers. 

Thus in this paper, the coordination number is defined as the number of \textit{molecules}, both water and ions of any type within the hydration shell at a snapshot in time, rather than the number of \textit{water molecules} within this radius. When averaged over time, this definition of the coordination number is equivalent to the integrated area under the radial distribution function (RDF) between the ion and all other molecules in solution up to the hydration shell radius, and the RDF can be measured by x-ray or neutron scattering (see Section~\ref{s:experimental} for mathematical definitions)~\cite{kameda_neutron_2018,dang_molecular_2006,varma_coordination_2006}. By considering all species within this hydration shell cutoff, we calculate the free energy to change the transporting volume for a given ion. The coordination shell thus includes the ion, coordinating waters, and coordinating counterions. We detail some other possible definitions of the de-coordination free energy in the Supplementary Materials Section~\ref{s:SI_dehydration_definition}.

In this study, we thus focus on the free energy of de-coordination in bulk solution, and we discuss its implications for ion movement within RO and NF membranes. We examine how free energy changes as a function of the coordination number, since the hydration shell formed by coordinated water molecules is commonly used to model the sterically-excluded size. In the process, we develop an approach to calculate the free energy associated with ion de-coordination as a function of the discrete coordination number states. Specifically, we determine trends in de-coordination free energies at elevated pressure and salinity. We look at these trends for ions relevant for high salinity brine treatment -- monovalent cations, divalent cations, and monovalent anions. We propose physical explanations for these trends, using evidence from the coordination shell structure and ion pairing events. 

We also investigate alternative ways to describe the complex geometrical constraints imposed by extreme confinement within polymer membranes, beyond simple spherical hydration shells. In Figure~\ref{fig:dehydration_visualization}, we visualize one hypothetical geometric distortion as the hydration shell (black dotted line) must go from a spherical shell to an ovular shell upon dehydration. In particular, we examine the free energy as a function of these geometric constraints, as one can argue this is a better description of the size of a hydrated ion traversing constrictions in the membrane. Because non-polarizable water models are not as physically accurate with ion-water interactions, we also estimate the sensitivity of our conclusions to choices of ion and water parameters. Finally, we discuss how our results deepen our understanding of ion transport in polymer membranes.

\section{Experimental} \label{s:experimental}

\subsection{Unbiased molecular dynamics simulations}

We ran molecular dynamics (MD) simulations for each pressure, concentration, and salt configuration to prepare initial configurations and parameters for umbrella sampling simulations. We packed water and ions into a box with Packmol~\cite{martinez_packmol_2009} at the appropriate ratio for the desired concentration. We then performed energy minimization, equilibration, and production MD simulations in Gromacs 2023.1 without any added bias~\cite{markidis_tackling_2015,abraham_gromacs_2015}. We used steepest descent energy minimization. Equilibration consisted of 50~ps in the canonical ensemble (NVT) at 300~K followed by 1~ns in the isobaric-isothermal ensemble (NPT) at 300~K and the desired pressure. Then we ran a production simulation in NPT for 20~ns at 300~K and the desired pressure. Input files for all simulations are provided at Github \url{https://github.com/shirtsgroup/solvation_shells}. 

We tested different water models and ion force field parameters in order to understand the robustness of our results. The water models we tested are TIP3P, TIP3P-FB, OPC3, TIP4P-EW, TIP4P-FB, OPC, and GOPAL. These choices include a range of three- and four-point water models that have been used for studies of electrolyte solutions and nanoscale confinement effects. TIP3P is a fast and common water model, albeit with known issues~\cite{jorgensen_comparison_1983}. TIP3P-FB is a reparameterized version of TIP3P that corrects some issues using the ForceBalance approach~\cite{wang_building_2014}. OPC3 is another three-point water model that has been optimized to better reproduce the properties of bulk water~\cite{izadi_accuracy_2016}. TIP4P-EW is a reparameterized version of the commonly used four-point water model TIP4P to correct issues that arise from Ewald summation of long-range electrostatics~\cite{horn_development_2004}. TIP4P-FB is a reparameterized version of TIP4P using the ForceBalance approach~\cite{wang_building_2014}. OPC is an optimized four-point water model~\cite{izadi_building_2014}. Finally, GOPAL is a four-point water model parameterized to get more accurate water properties over a wide range of both temperature and pressure~\cite{paliwal_water_2023}. Since non-polarizable water models are less physically accurate than well-parameterized polarizable water models when used in electrolyte solutions, we performed a thorough study of non-polarizable models and discuss the sensitivity of our results to the choice of water model (Section~\ref{s:force_field}), with the expectation that findings that are independent of water model are more likely to be true physical results. When possible, we used ion parameters optimized for the given water models, except for the newly developed GOPAL model. Li et al.~optimized parameters for TIP3P and TIP4P-EW in 2015~\cite{li_systematic_2015}. Li et al.~optimized parameters for divalent ions with OPC3, OPC, TIP3P-FB, and TIP4P-FB in 2020~\cite{li_systematic_2020}, and Sengupta et al.~optimized parameters for monovalent ions with these water models in 2021~\cite{sengupta_parameterization_2021}. For GOPAL, we used the ion parameters optimized for OPC as the water model most closely resembles OPC. 

\subsection{Definition of the RDF and hydration shell radius}

The RDF provides a description of the particle density radially from a reference and is commonly used to examine local molecular environments. We adopt the typical definition of the hydration shell radius as the distance that marks the change from the first shell of coordinating waters to the second shell, which is determined by the first minimum in the ion-water RDF. We calculate the ion-water RDF by Equation~\ref{eq:RDF} for distances between the ions and oxygen atoms in the water molecules in the simulations~\cite{gowers_mdanalysis_2016}, with examples of the determination of the hydration shell radius for Na$^+$ and Cl$^-$ shown in Figure~\ref{fig:hydration_radius}.  Then the RDF is:

\begin{equation}
    g_{ab}(r) = \frac{1}{N_a} \frac{1}{(N_b / V)} \sum_i^{N_a} \sum_j^{N_b} \left< \delta \left( |\textbf{r}_i - \textbf{r}_j| - \text{r} \right) \right>
    \label{eq:RDF}
\end{equation}

\noindent where $g_{ab}(r)$ is the RDF between group $a$ and $b$, $N_a$ is the number of particles in group $a$, $V$ is the volume of the system, and $\bf{r_i}$ is the vector coordinates of particle $i$. 

\subsection{Umbrella sampling in ion coordination number}

\begin{figure*}[ht!]
    \centering
    \includegraphics[width=\textwidth]{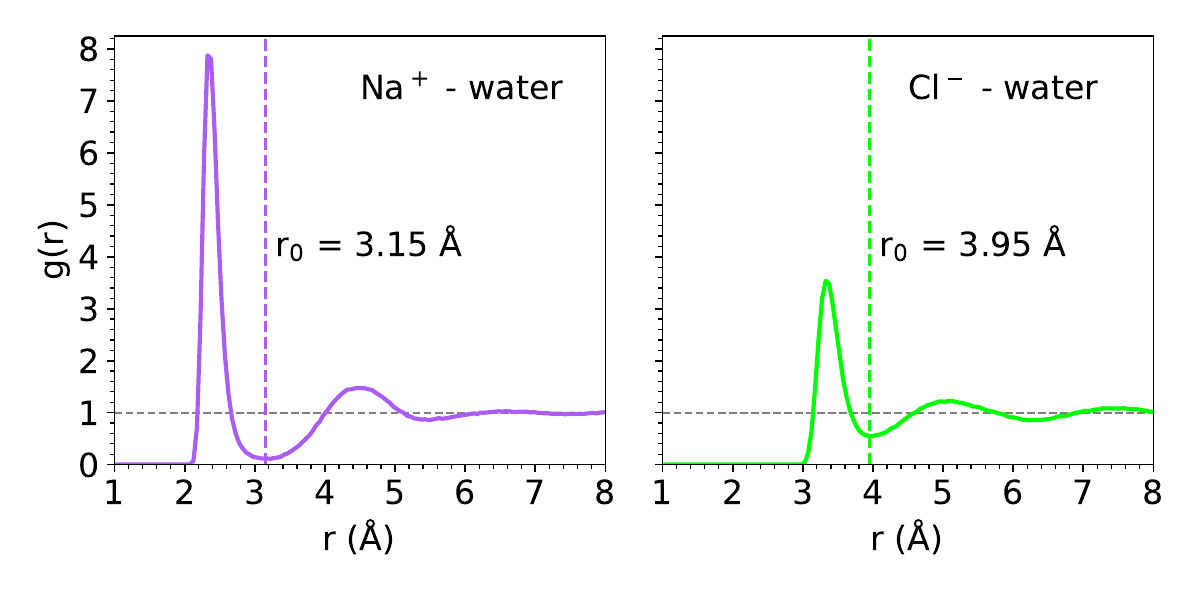}
    \caption{Hydration shell radius is determined by the first minimum in the ion-water RDFs, shown here for Na$^+$ and Cl$^-$. The cutoff (r$_0$) for Na$^+$ is 3.15~\AA, and the cutoff for Cl$^-$ is 3.95~\AA. The ion-water oxygen RDFs were generated from a 20~ns simulation of NaCl solution at infinite dilution and 1~bar using OPC3 water and optimized ion parameters~\cite{sengupta_parameterization_2021}.}
    \label{fig:hydration_radius}
\end{figure*}

We performed umbrella sampling simulations to calculate free energy surfaces (FES) in the ion coordination number and the ion de-coordination free energies. We used Gromacs 2023.1 patched with PLUMED 2.10.0-dev to run these biased simulations~\cite{the_plumed_consortium_promoting_2019}. We applied a harmonic restraint to the coordination number of a single ion. To use the ion coordination number as the collective variable requires a switching function to make the discrete coordination number a continuous, differentiable function of system coordinates at each time step. We used the switching function from Brancato \& Barone (Eq.~\ref{eq:switching_function}) since it has been shown to work well for the ion coordination number collective variable, and it has only one tunable parameter, $a$~\cite{brancato_free_2011}. 
\begin{equation}
    s = \sum_{i=1}^{N} \left [ 1 - \frac{1}{ 1 + \exp{\left ( -a (r_i - r_0) \right )} } \right ]
    \label{eq:switching_function}
\end{equation}
\noindent $N$ is the number of all other molecules in the system, $r_i$ is the distance between the biased ion and molecule $i$, and $r_0$ is the spherical hydration shell cutoff inside which the coordination number is desired. We implemented this function using a modified version of the \texttt{Q} function in PLUMED~2 (Eq.~\ref{eq:Q_function}). 
\begin{equation}
    s = \sum_{i}^{N} \left [ \frac{1}{ 1 + \exp{\left ( \beta (r_i - \lambda r_0) \right )} } \right ]
    \label{eq:Q_function}
\end{equation}
\noindent To conform to the switching function from Brancato \& Barone, we set $\beta = -a$ and $\lambda = 1$, and we subtracted the output from the number of molecules $N$. We determined the hydration shell cutoff $r_0$ from ion-water RDFs calculated from the unbiased MD simulations. We used the MDAnalysis Toolkit~\cite{gowers_mdanalysis_2016,michaudagrawal_mdanalysis_2011} \texttt{SolvationAnalysis}~\cite{cohen_solvationanalysis_2023} with a \texttt{scipy.find\_peaks} wrapper to locate the first minimum in the ion-water RDF. This algorithm uses peak properties and simple comparisons of neighboring values to determine the locations of local minima. It was robust enough to give consistent hydration shell cutoffs for all conditions. Figure~\ref{fig:hydration_radius} shows the ion-water RDF and the hydration shell radius determined for Na$^+$ and Cl$^-$. Notably, the RDF for chlorine shows significant water density near the cutoff, unlike Na$^+$ where the radial water density is near zero at the minimum.

We tuned the switching function parameter $a$ by running short (10~ns) umbrella sampling simulations with the continuous coordination number bias, constructing the free energy surfaces from these biased simulations, and qualitatively comparing them. We expected that tuned parameters should give a smooth free energy surface as a function of coordination number and show a minimum-energy coordination number similar to the coordination number determined in the unbiased MD simulations. We visually balanced these two aspects of the free energy surface in order to determine the value of the switching function parameter. 

Importantly, for the final de-coordination free energies, we only calculated the free energies for the discrete coordination number states, rather than the continuous coordination numbers calculated by the switching function. The continuous coordination number is strongly dependent on the choice of the switching function parameter. Molecules near the cutoff can contribute partial coordination numbers, which artificially inflate the sampling of these configurations. We provide examples of this behavior in the Supplementary Materials Section~\ref{s:SI_switching_function}. However, by calculating the free energies of the discrete coordination number states by reweighting the umbrella simulations generated with the continuous coordination number function, we largely eliminated the dependence of the free energy on the switching function parameter as shown in Supplementary Materials Section~\ref{s:SI_switching_function}. Therefore, the switching function only needs to be tuned enough so that the discrete coordination number configurations of interest are sufficiently sampled.

We ran short (100~ps) umbrella simulations starting from the final frame of our unbiased MD simulations, incrementally spanning the collective variable space of interest. We used these short simulations to generate initial configurations in order to avoid the large forces introduced when biasing the final frame of the unbiased MD to the umbrella centers. We ran these short umbrella simulations with the umbrella sampling parameters discussed below. 

To calculate the free energies as a function of coordination number, we ran each umbrella simulation, initialized as described above, for 100~ns in NPT at 300~K and the desired pressure. The 100~ns simulations ensured there was sufficient sampling even for highly unfavorable regions of coordination number space. All simulations were run with v-rescale temperature coupling (\texttt{tau\_t} = 1~ps) and Parrinello-Rahman pressure coupling (\texttt{tau\_p} = 10~ps). We applied harmonic restraints at umbrella centers spanning the relevant coordination number space for each ion. The force constants for the individual harmonic restraints were chosen to ensure sufficient overlap between neighboring umbrellas. Most configurations of concentration, pressure, and ion used 16 evenly-spaced umbrella centers with equal force constants; however, specific umbrella centers, force constants, and switching function parameters are included in the Supplementary Materials Section~\ref{s:SI_umbrella_configs}. In general, we looked at regions of the coordination number space with barriers less than 100~kJ/mol, since we expect mechanisms with higher free energy barriers to be much less relevant for RO and NF membrane separations~\cite{shefer_applying_2022,richards_experimental_2013}. We note that the actual barriers to membrane transport are lower, since interactions with a membrane would partially compensate for ion dehydration.

We estimated the free energy as a function of the coordination number from the biased simulations using the Multistate Bennett Acceptance Ratio (MBAR) implemented in \texttt{pymbar}~\cite{shirts_statistically_2008}. We calculated the discrete coordination number as the number of molecules within the hydration shell radius at each time step. We reweighted the configurations in the continuous coordination number to the discrete coordination number distribution. We then calculated the free energy for each discrete coordination number state ($\Delta G(\text{CN})$) as given in Equation~\ref{eq:discrete_FE}, which corresponds to the free energy to go from the minimum-energy coordination number to coordination number CN. 

\begin{equation}{
    \Delta G(\text{CN}}) = -k_B T \exp{\left( \sum_k^K \sum_n^{N_k} w_{n,k} \, \delta (\text{CN}_{n,k} - \text{CN}) \right)}
    \label{eq:discrete_FE}
\end{equation}

\noindent In Equation~\ref{eq:discrete_FE}, $k_B$ is the Boltzmann constant, $T$ is temperature, $K$ is the number of umbrella simulations, and $N_k$ is the number of samples in umbrella $k$. The weights calculated from the biased simulations $w_{n,k}$ are summed for each coordination number CN, where $\text{CN}_{n,k}$ is the discrete coordination number for configuration $(n,k)$. We calculated the weights for sample $n$ in umbrella simulation $i$ using Equation~\ref{eq:MBAR_weights} with the bias applied to the continuous coordination number.
\begin{equation}
    w_{n,i} = \frac{\exp{( f_i- u_i(\bf{x_n}) )}}{ \sum_{k=1}^{K}{N_k \exp{(f_k - u_k(\bf{x_n}) )} } }
    \label{eq:MBAR_weights}
\end{equation}

\noindent $f_i$ is the reduced free energy $\left( \frac{G}{k_B T} \right)$ for applying umbrella simulation $i$, estimated from the bias via MBAR~\cite{shirts_statistically_2008}. $u_i(\bf{x})$ is the reduced potential $\left( \frac{U}{k_B T} \right)$ of the umbrella biasing potential $i$ for configuration $\bf{x}$. $\bf{x_n}$ are the atomic coordinates for sample $n$.  We estimated the uncertainty in the free energies of the discrete coordination number states with 50 bootstrapped sets of samples from the $K$ umbrella simulations.

Therefore, the de-coordination free energies are the sequential differences from the minimum-energy coordination number to lower coordination numbers. Our methodology reweighting simulations biased in a continuous variable to discrete distributions is a novel and nontrivial procedure with applications for other discrete collective variables, such as hydrogen bonds or contacts in proteins. Example calculations where we reweight configurations from the umbrella simulation to the discrete coordination number states are available on Github at \url{https://github.com/schwinns/solvation_shells}.

\subsection{Analysis of hydration shell structure}

We quantified the limiting size of the hydration shell by looking at the cross-sectional areas of convex hulls formed by molecules in the hydration shell. These dimensions are the least restrictive dimensions if the hydration shell were to move through a constriction in a membrane. For each frame, we created a convex hull from the van der Waals volume of atoms within the hydration shell. We then determined the principal axis of each polyhedron and calculated the largest cross-sectional area along the principal axis. We calculated the convex hull using \texttt{scipy.spatial.ConvexHull}, a Python interface for the Qhull library. We used \texttt{sklearn.decomposition.PCA} from \texttt{scikit-learn} to perform the principal component analysis on the convex hull vertices. More detailed information on how we calculated the shell limiting area is included in Supplementary Materials Section~\ref{s:SI_limiting_area} with example code at \url{https://github.com/shirtsgroup/solvation_shells/}. 

With these cross-sectional areas calculated for each configuration, we estimated the free energy to change the cross-sectional area. We calculated the free energy surface in the cross-sectional area using the same umbrella sampling simulations biased in the continuous coordination number. Rather than reweighting to a discrete distribution as in the case of the coordination number free energies, we reweighted the configurations in the continuous coordination number to the (continuous) cross-sectional area distribution.

\section{Results}

\subsection{Elevated operating pressures do not significantly affect ion de-coordination}

We first examined how the ion de-coordination free energy changes at high pressure in order to understand whether changes in the hydration shell contribute to the decreased water-salt selectivity observed during brine treatment~\cite{davenport_high-pressure_2018,chen_steric_2023}. We tested two pressure conditions near the extremes of membrane operating pressures, specifically 1~bar and 150~bar. We studied the effect of system pressure on the shell rather than a pressure gradient since the pressure gradient across an individual coordination shell would be small in membrane applications, even at elevated pressure needed for brine treatment.

\begin{figure}[H]
    \centering
    \includegraphics[width=0.9\textwidth]{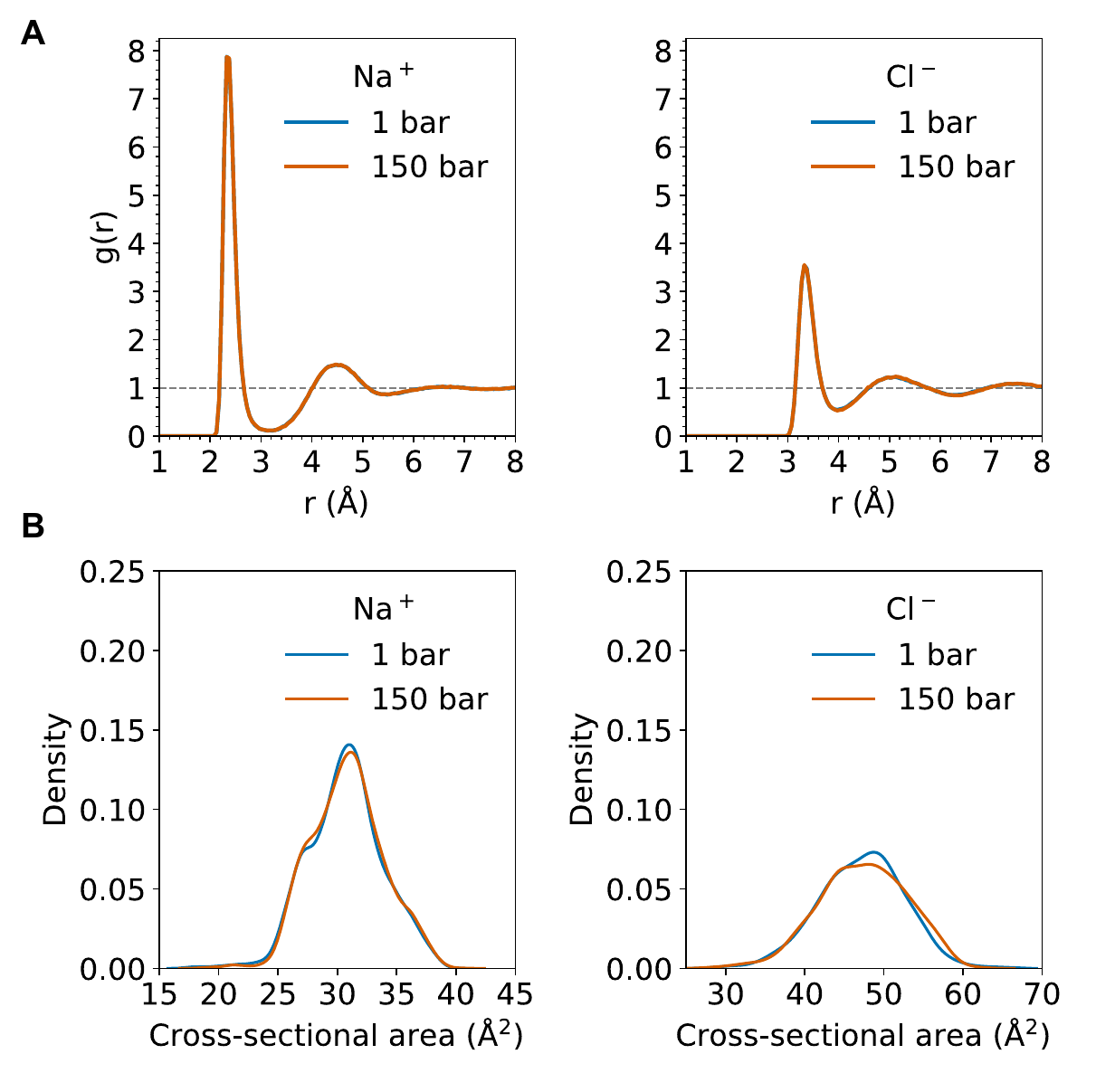}
    \caption{Coordination shell size and structure do not change with increasing pressure. (\textbf{A}) The ion-water RDFs for Na$^+$ and Cl$^-$ at 150~bar are indistinguishable from those at 1~bar. The low pressure condition (blue) overlaps nearly exactly with the high pressure condition (orange). (\textbf{B}) Distributions of the limiting cross-sectional areas of the coordination shells do not change significantly at elevated pressure. Distributions are shown using kernel density estimations with Scott's Rule to determine the bandwidth~\cite{scott_multivariate_1992}. Shell structure data were calculated from 20~ns trajectories at infinite dilution. Other ions showing similar distributions are shown in Supplementary Materials Section~\ref{s:SI_complete_pressure}.}
    \label{fig:structure_pressure}
\end{figure}

We find that high pressure does not noticeably change the coordination shell structure in solution, although we observed some small differences for anions as noted below. We quantified the size and shape of the coordination shells of all ions in unbiased MD simulations at both 1~bar and 150~bar. The radial structure does not change for either cations or anions, which is expected for pressures in this range since aqueous solutions have low compressibility~\cite{zhang_dissolving_2022,skinner_structure_2016}. However, significant changes in RDFs have been reported at very high pressures~\cite{tonti_how_2021}. Figure~\ref{fig:structure_pressure}A shows the ion-water RDFs for sodium and chlorine ions at infinite dilution, and the high pressure RDF is indistinguishable from the low pressure condition. The ion-water peak and the hydration shell radius remain unchanged. Similarly, the distributions of largest cross-sectional areas along the principal axis of the coordination shell are mostly independent of pressure. For anions, the increased pressure slightly increases the spread of the distribution, which could indicate some small stabilization of the coordination shell. We show results for Na$^+$ and Cl$^-$ for clarity, since the trends for these ions are representative of the cations and anions we examined. We explicitly acknowledge some small but notable exceptions with complete data presented in the Supplementary Materials. RDFs, limiting area distributions, and free energies of coordination number states for all ions as a function of pressure are in Supplementary Materials Section~\ref{s:SI_complete_pressure}.

\begin{figure}[H]
    \centering
    \includegraphics[width=0.75\textwidth]{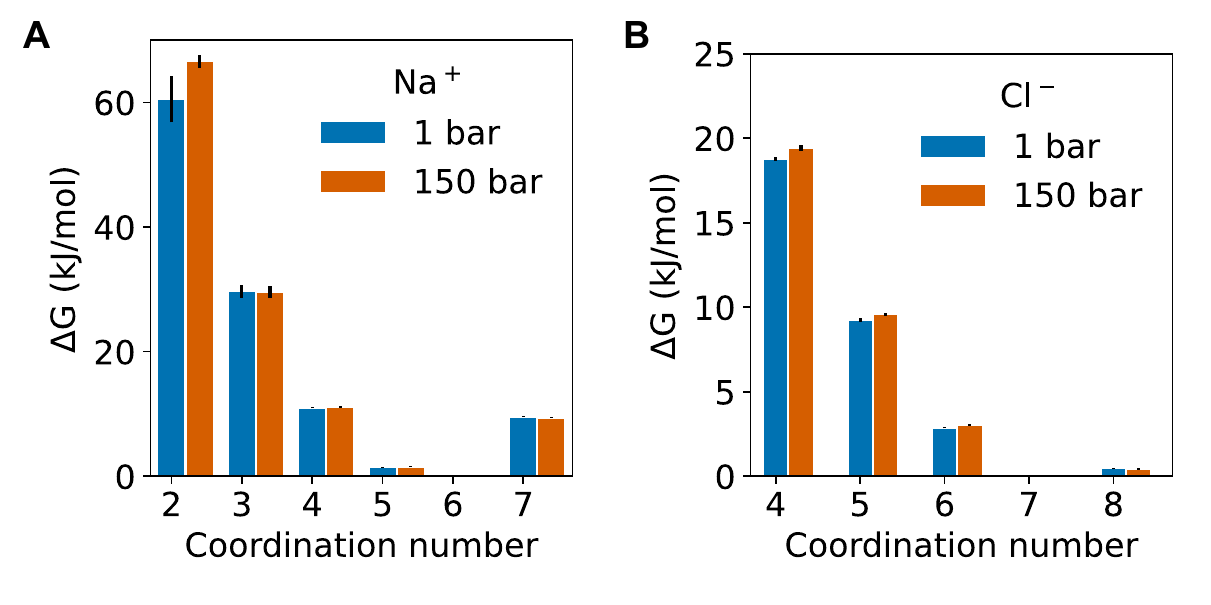}
    \caption{Free energies of coordination number states are not strongly dependent on pressure. (\textbf{A}) The de-coordination free energies for Na$^+$ at infinite dilution do not show a statistically significant trend at increasing pressure. (\textbf{B}) The de-coordination free energies for Cl$^-$ at infinite dilution show a slight increase with increasing pressure.}
    \label{fig:free_energy_pressure}
\end{figure}

High pressure does not change the de-coordination free energies for cations, but it causes a small, systematic increase in the de-coordination free energies for anions. A few cations (Li$^+$, Ca$^{2+}$, and Sr$^{2+}$) also show small changes in de-coordination free energies at elevated pressure. The free energy to reach coordination number states below the minimum-energy coordination number are statistically indistinguishable for Na$^+$ as shown in Figure~\ref{fig:free_energy_pressure}A, shown as a typical case for cations. The smaller, more tightly coordinated Li$^+$ is less stable at high pressure by about 6~kJ/mol per de-coordinated molecule. The solvation shells of the large divalent cations Ca$^{2+}$ and Sr$^{2+}$ are slightly more stable at high pressure, especially at low coordination number (Figure~\ref{fig:SI_all_fe_pressure}. The solvation shells of all anions tested are also slightly more stable at higher pressure, as shown for Cl$^-$ in Figure~\ref{fig:free_energy_pressure}B. The increase in free energy for anions at elevated pressure is consistent across the anions tested, but the changes are likely negligible ($< 1$~kJ/mol). Larger ionic radii appear to correlate with a higher de-coordination free energy at elevated pressure.

\subsection{High salinity decreases cation de-coordination free energies}

\begin{figure}[H]
    \centering
    \includegraphics[width=0.75\textwidth]{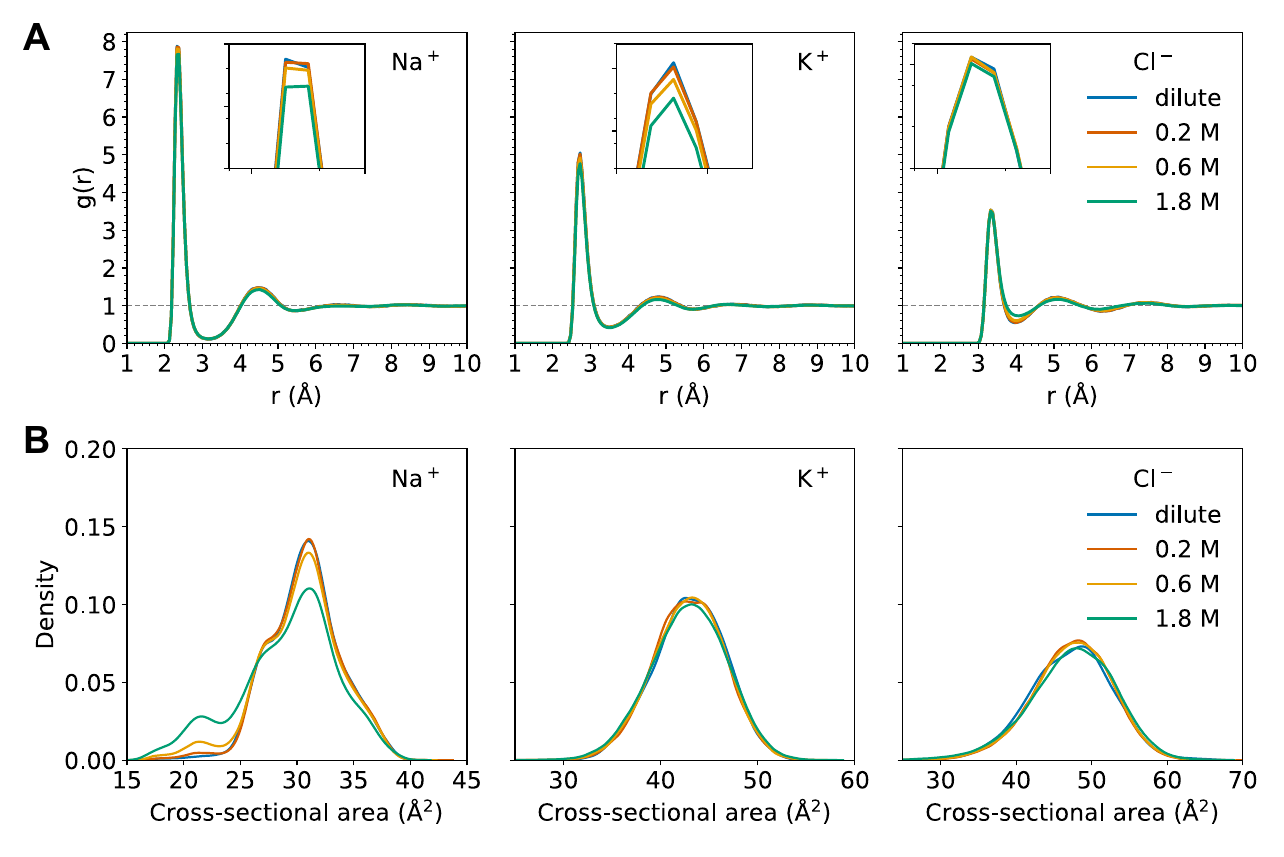}
    \caption{Coordination shell structure analyses show more small cross-sectional area configurations at increasing salinity for Na$^+$ but not K$^+$ or Cl$^-$. (\textbf{A}) The ion-water RDFs for Na$^+$ and K$^+$ show smaller first-peak magnitudes at higher concentrations. For Cl$^-$, higher concentrations increase the density of water molecules near the hydration shell cutoff. The tick marks on the inset image are the same as the main axes (i.e. major ticks are at integers and minor ticks are at 0.2 increments).(\textbf{B}) Distributions of the limiting cross-sectional areas of the Na$^+$ coordination shells show more low area configurations at high salinity, which often correspond to configurations with ion pairing events. The limiting cross-sectional areas of K$^+$ and Cl$^-$ shells do not change at high salinity. Shell structure data were calculated from 20~ns trajectories at 1~bar. Other ions showing similar distributions are shown in Supplementary Materials Section~\ref{s:SI_complete_concentration}.}
    \label{fig:structure_concentration}
\end{figure}

High salinity slightly reduces the size of the cation coordination shell but does not change the size of the anion coordination shell. The first ion-oxygen peak in the RDF (Figure~\ref{fig:structure_concentration}A) decreases in magnitude for most cations, which corresponds to more low coordination number configurations. For Ca$^{2+}$ and Sr$^{2+}$, the magnitude of the first peak increases at high concentration as shown in Supplementary Materials Figure~\ref{fig:SI_all_RDFs_concentration}B and C. However, the hydration shell cutoff remains the same at increased concentration for all cations. With the exceptions of Na$^+$ and Mg$^{2+}$, the distribution of restrictive cross-sectional areas does not change with elevated salinity as shown for K$^+$ in Figure~\ref{fig:structure_concentration}B. However, for Na$^+$ and Mg$^{2+}$, the increased frequency of low coordination number states causes a significant peak at low cross-sectional area, capturing some of the discrete coordination number changes. This discreteness is likely due to the combination of small ionic size and high charge density. We thus expect these shells to be more rigid than those of other ions. For anions, the density of molecules near the hydration shell radius increases but does not change where the shell cutoff occurs. The less structured minimum indicates that the anions are disrupting the water structure. The limiting area of the coordination shell---the maximum cross-sectional area along the principal axis---does not change with the increased density near the cutoff, indicating that the molecules near the cutoff are not as important for reducing this restrictive size of the coordination shell as the tightly coordinated molecules. Figure~\ref{fig:structure_concentration}B shows the distribution of the cross-sectional areas at increasing salinity, and for Cl$^-$, the distributions are unchanged.

\begin{figure}[H]
    \centering
    \includegraphics[width=\textwidth]{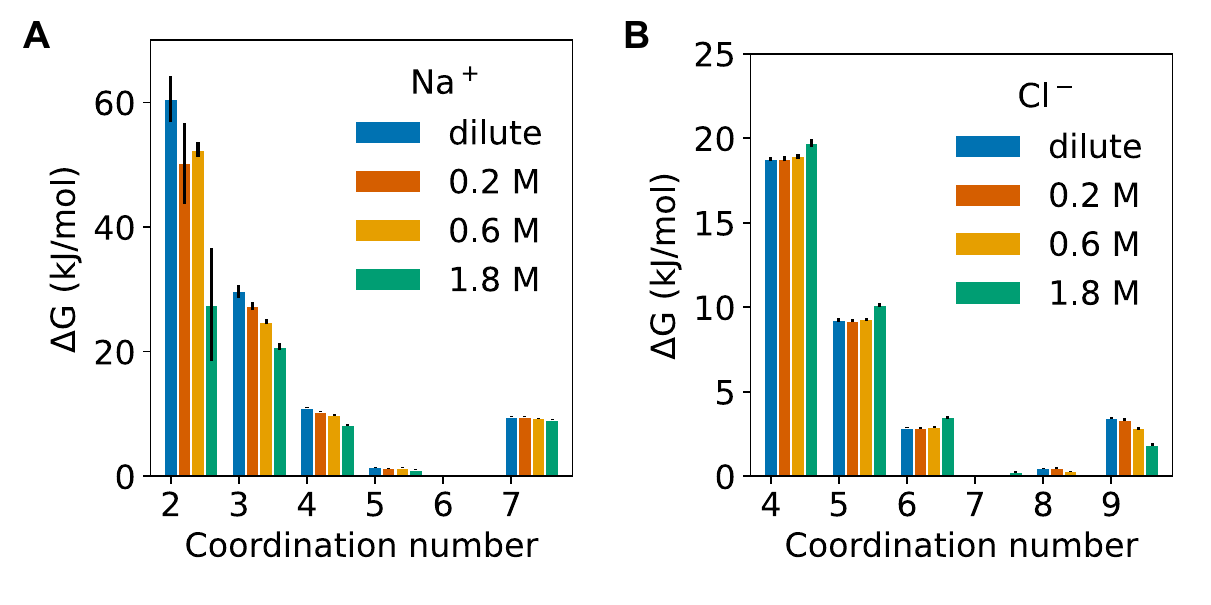}
    \caption{Free energies of coordination number states decrease with increasing salinity for cations. (\textbf{A}) The de-coordination free energies for Na$^+$ decrease with increasing concentration. (\textbf{B}) The de-coordination free energies for Cl$^-$ only change at 1.8~M, where the minimum-energy coordination number increases to 8.}
    \label{fig:free_energy_concentration}
\end{figure}

Similar to the effect of salinity on size, we find that high concentrations decrease cation de-coordination free energies but have little effect on the anion de-coordination free energies. The free energies associated with low coordination number states for cations are substantially lower at high salinity as shown in Figure~\ref{fig:free_energy_concentration}A for Na$^+$, which indicates that cations more readily de-coordinate when there are many ions nearby. This conclusion is consistent with the results from unbiased MD at high concentration showing more frequent low coordination number configurations (Figure~\ref{fig:structure_concentration}). We do not observe this trend for Mg$^{2+}$. The de-coordination energies for Mg$^{2+}$ do not show significant differences with increasing salinity, likely due to its strong electrostatic interactions and small ionic size. Concentration dependence results, including RDFs, limiting area distributions, and free energies, for Mg$^{2+}$ and all other ions tested are in Supplementary Materials Section~\ref{s:SI_complete_concentration}. Additionally, Figure~\ref{fig:all_ions_pressure_concentration}B summarizes the de-coordination free energy results for the first two de-coordinations of all ions. For anions, the only difference in de-coordination free energies with increasing concentration occurs at the highest concentration tested. At 1.8~M, the minimum-energy coordination number increases. In Figure~\ref{fig:free_energy_concentration}B, the minimum-energy coordination number for Cl$^-$ increases from 7 to 8 at 1.8~M, suggesting that, on average, nearby ions force one more molecule into the shell. The additional de-coordination thus requires more energy to reach low coordination number states. 

\subsection{Elevated pressure and salinity could improve ion-ion selectivity between some ions}

Partial de-coordination free energies follow the same ion size and valency trends seen in experimental hydration free energies; larger bare ions are easier to de-coordinate and higher valency ions are more strongly coordinated~\cite{sengupta_parameterization_2021, li_systematic_2020, schmid_new_2000, marcus_thermodynamics_1991}. However, we find deviations from these patterns for the first two de-coordinations. For example, trends in hydration free energies predict that divalent ions are approximately 4x more strongly coordinated than similarly sized monovalent ions, but the first and second de-coordinations are only approximately 2x more strongly coordinated. 

\begin{figure}[H]
    \centering
    \includegraphics[width=\textwidth]{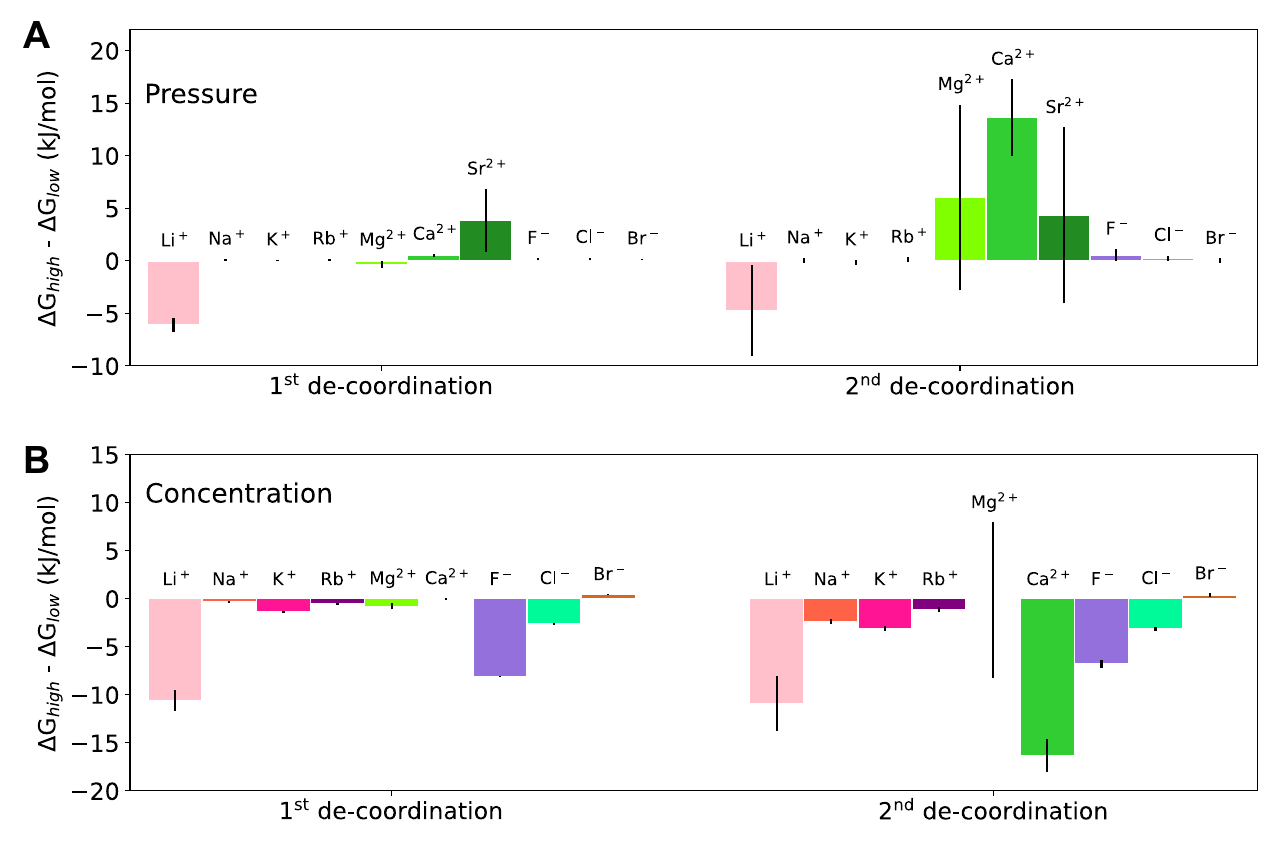}
    \caption{The first two de-coordination free energies could provide insight into differences in ion-ion selectivity at high pressure and salinity. De-coordination free energies are calculated as the free energy differences between sequential coordination number states. For example, the first de-coordination free energy for Na$^+$ is the free energy difference between coordination number 6 and coordination number 5. The second de-coordination free energy is the difference between 5 and 4. Here, we show the differences between the highest and lowest pressure and concentration conditions for the first two de-coordination free energies. A positive $\Delta G_{high} - \Delta G_{low}$ indicates that de-coordination is more difficult at the elevated condition. (\textbf{A}) The free energy change between high and low pressure for the first and second de-coordinations for all ions at infinite dilution. Ions tested show no change in the de-coordination free energy at high pressure, except for Li$^+$, Ca$^{2+}$, and Sr$^{2+}$. For these ions, high pressure could improve selectivity. (\textbf{B}) We show the free energy change between high and low concentration for the first and second de-coordinations for all ions at 1~bar. Relative ion de-coordination free energies are maintained at high concentration. The first two de-coordination free energies for Li$^+$, F$^-$, and Ca$^{2+}$ decrease by more than 5~kJ/mol at high concentration.}
    \label{fig:all_ions_pressure_concentration}
\end{figure}

We compare the trends in ion de-coordination free energies observed at low pressure and low salinity to those calculated for high pressure and high salinity. Our results in Figure~\ref{fig:all_ions_pressure_concentration} suggest that high pressure and salinity could potentially be used to improve ion-ion selectivity for some ions, if de-coordination can be leveraged as the governing mechanism. Figure~\ref{fig:all_ions_pressure_concentration}A shows the difference in the de-coordination free energy between low and high pressure for all ions tested. At high pressure, the first and second de-coordination free energies only change significantly for Li$^+$, Ca$^{2+}$, and Sr$^{2+}$. Therefore, increasing pressure could improve ion-ion selectivity between these ions and those that are unchanged at high pressure. For example by adding the two free energy differences for a given ion in Figure~\ref{fig:all_ions_pressure_concentration}A, the barrier for Li$^+$ to de-coordinate twice decreases by 10.8~kJ/mol at 150~bar, while the free energy for Mg$^{2+}$ to de-coordinate twice changes negligibly. As a result, high pressure theoretically would increase the de-coordination-controlled Li$^+$/Mg$^{2+}$ selectivity. Figure~\ref{fig:all_ions_pressure_concentration}B shows the difference in the de-coordination free energy between low and high salinity for all ions tested. At high concentration, both of the first two de-coordination free energies decrease by more than 5~kJ/mol for Li$^+$, F$^-$, and Ca$^{2+}$. Therefore, selectivity towards these ions could be higher at elevated salinity, since they will more easily de-coordinate and thus more easily traverse membrane constrictions. 

\subsection{The free energy to change the restrictive cross-sectional area is not dependent on ion concentration}

\begin{figure}[H]
    \centering
    \includegraphics[width=\textwidth]{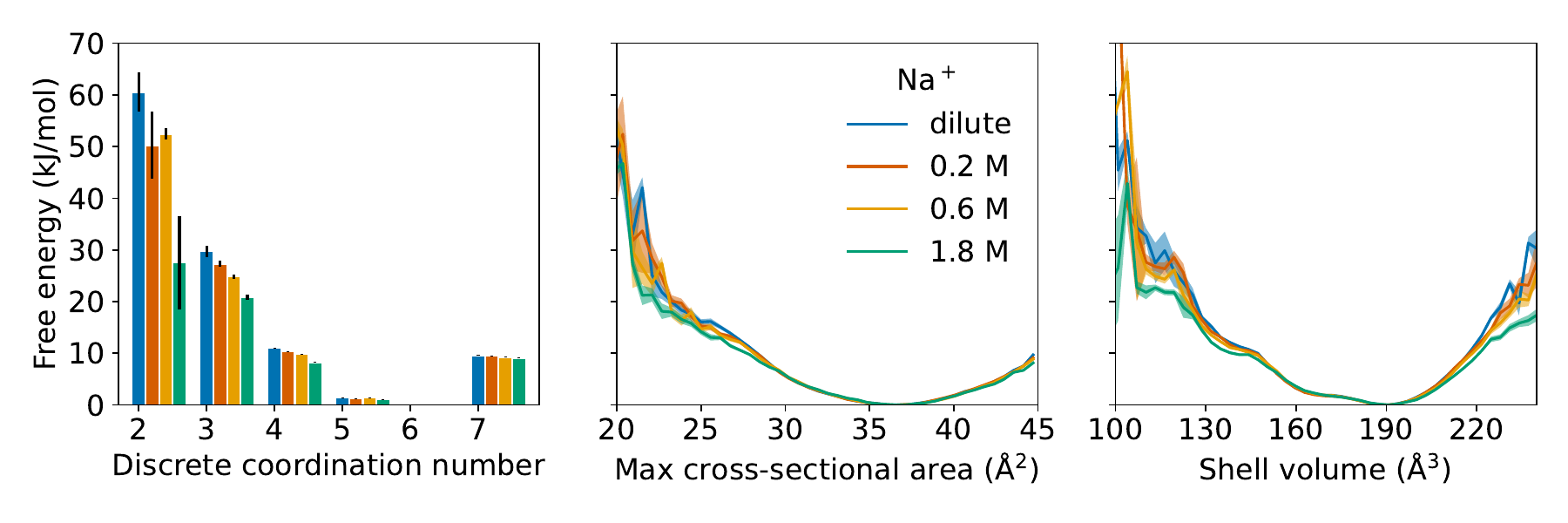}
    \caption{The free energy surfaces in limiting cross-sectional area of the shell and in shell volume show little dependence on concentration for Na$^+$ at 1~bar, in contrast to the free energies in coordination number. In the three panels, we compare the free energies of the discrete coordination number states, the maximum cross-sectional area of the shell, and the shell volume on the same scale. The free energy surface as a function of the maximum cross-sectional area along the principal axis of the coordination shell does not show statistically significant differences at elevated salinity, unless the shell can reach very low cross-sectional areas. The free energy surface in shell volume shows a similar trend, but the volume collective variable captures the discreteness of coordination number better than the cross-sectional area.}
    \label{fig:free_energy_in_area_volume}
\end{figure}

We quantified the restrictive, limiting dimension to transport as the maximum cross-sectional area along the principal axis of the occupied shell volume. We characterized how this critical size changes as an ion de-coordinates at a range of pressures and concentrations. To more specifically calculate the free energy penalty, we estimated the free energy surface as a function of the (continuous) cross-sectional area of the shell. Similarly to the free energies of the discrete coordination number states, we reweighted the configurations in the continuous coordination number to the cross-sectional area distribution. We argue that the cross-sectional area is more important than coordination number or shell volume for an ion moving through a constriction, but we quantified the free energies in all three collective variables as shown in Figure~\ref{fig:free_energy_in_area_volume}. 

For moderate reduction in hydrated size, it is easier to reduce the coordination number than to reduce the free energy the maximum cross-sectional area. For example, for Na$^+$ at 1~bar and infinite dilution, the free energy to reduce the coordination number by 1/3 (i.e. from 6 to 4) is 11.0~$\pm$~0.1~kJ/mol (Figure~\ref{fig:free_energy_in_area_volume} left panel), while the free energy to reduce the cross-sectional area by 1/3 (i.e. from 36~\AA$^2$ to 24~\AA$^2$) is 18.3~$\pm$~0.9~kJ/mol (Figure~\ref{fig:free_energy_in_area_volume} middle panel). However, the free energy to reduce the cross-sectional area to the average area for coordination number 4 (i.e. from 36~\AA$^2$ to 30~\AA$^2$) is 5.3~$\pm$~0.1~kJ/mol. Reducing the coordination number by 1/3 is not equivalent to reducing the cross-sectional area by 1/3; rather it is easier than reducing the cross-sectional area by 1/3.  

Notably, the free energy to change cross-sectional area does not change significantly at increased concentration, despite the strong affect of salinity on the free energies of coordination number states. However, reducing the cross-sectional area to the lowest sampled region ($\sim$20~\AA$^2$) is easier at high concentration as seen in Figure~\ref{fig:free_energy_in_area_volume}A. Calculating the free energy as a function of the shell volume shows a similar lack of concentration dependence, but the free energy surface captures some of the discreteness from the coordination number, not seen in the cross-sectional area. In Figure~\ref{fig:SI_correlations_area_volume_cn}, we show the correlation among the coordination number, the maximum cross-sectional area along the principal axis of the shell, and the shell volume. Many configurations with low cross-sectional area do not have low coordination number, indicating that simply rearranging the coordination shell could facilitate movement through a constriction without de-coordination.


\subsection{Well-optimized force field parameters yield qualitatively similar free energies}\label{s:force_field}

We examined how the choice of non-polarizable water model and ion parameters affected the de-coordination free energies as a test of the robustness of the trends we observed. Figures and further discussion are included in Supplementary Materials Section~\ref{s:SI_force_field}. We calculated the free energies for Na$^+$ at infinite dilution and 1~bar for the seven water models presented in the Methods. We also tested two sets of ion parameters with OPC3 water -- Li \& Merz 2015~\cite{li_systematic_2015} optimized for TIP3P and Sengupta et al. 2021~\cite{sengupta_parameterization_2021} optimized for OPC3. 

We find that trends in the de-coordination free energies are consistent across water models (Figure~\ref{fig:SI_water_models}B). Notably, the water model that differs most significantly - GOPAL - does not have optimized ion parameters. There are a few subtle differences; for example, TIP3P is slightly more strongly coordinated than the other three-point water models. The first and second de-coordination free energies are 1.6~kJ/mol and 1.0~kJ/mol higher for TIP3P than OPC3. Four-point water models are systematically more strongly coordinated than three-point water models. The first and second de-coordination free energies are 2.9~kJ/mol and 1.8~kJ/mol higher for OPC than OPC3. The de-coordination free energies are statistically indistinguishable between ion parameters as shown in Figure~\ref{fig:SI_ion_params}B. Both the Li \& Merz~2015 and the Sengupta et al.~2021 parameter sets were optimized to reproduce experimental hydration free energies with three-point water models. Therefore, it is unsurprising that the ion parameters perform similarly for our calculations. 

Aside from these few small differences, the overall lack of dependence on force field parameters suggests that the behavior we observe is physical, i.e. not an artifact of parameterization. Non-polarizable force fields have been shown to overestimate the charge density of ions in aqueous solution, which likely results in de-coordination free energies that are systematically overestimated as well~\cite{kirby_charge_2019}. Additionally, classical force fields do not model dynamic properties of aqueous electrolytes well~\cite{panagiotopoulos_dynamics_2023}. On the other hand, RDFs and other structural properties are typically well-reproduced with these force fields. Therefore, we acknowledge that the exact numerical values of the barriers we present may not be as accurate as higher levels of theory, but we conclude that the magnitudes and trends are representative of ions in solution.


\section{Discussion}

\subsection{Geometric description of the ion de-coordination process}

In this study, we carefully defined the process of reducing the effective size of hydrated ions as de-coordination. De-coordination corresponds to removing any coordinated molecule from the coordination shell. This definition helps clarify the perspective on how ions move through membrane constrictions. When considering such motion, we must consider multi-species, non-spherical complexes rather than spherical shells of tightly coordinated water molecules, especially in the high salinity environments necessary for brine treatment. 

To further quantify the important geometric considerations of nonspherical molecular clusters, we calculated the maximum cross-sectional area along the principal axis of the occupied shell volume. This alternative measure of hydrated ion size captures asymmetry and potentially can capture low energy configurations introduced during de-coordination and not described by coordination number.

In Figure~\ref{fig:free_energy_concentration}, we showed that the free energy to de-coordinate significantly decreases at high concentration for cations, while it does not change for anions. However, the significant decrease in free energy at high concentration does not hold for cross-sectional area, as shown in Figure~\ref{fig:free_energy_in_area_volume}. We hypothesize that high concentration decreases the barrier by rearranging the loosely coordinated molecules, which are farther from the ion. These molecules will not contribute much to the restrictive cross-sectional area, since we calculate the restrictive cross-sectional area as the area \textit{perpendicular} to the longest axis rather than \textit{parallel} to the longest axis. Therefore, the lower barrier to de-coordinate at high salinity may not necessarily correspond to a lower barrier to move through a constriction in the membrane. Comparing hydration free energies, or even de-coordination free energies, may not fully capture the important barriers associated with the ion de-coordination mechanism. The penalty to reduce the hydrated ion size often involves de-coordination but may only require changing the restrictive cross-sectional area, which our results suggest is a smaller barrier than the barrier associated with de-coordination. 

\subsection{Physical explanations for observed trends}

\begin{figure}[H]
    \centering
    \includegraphics[width=0.8\textwidth]{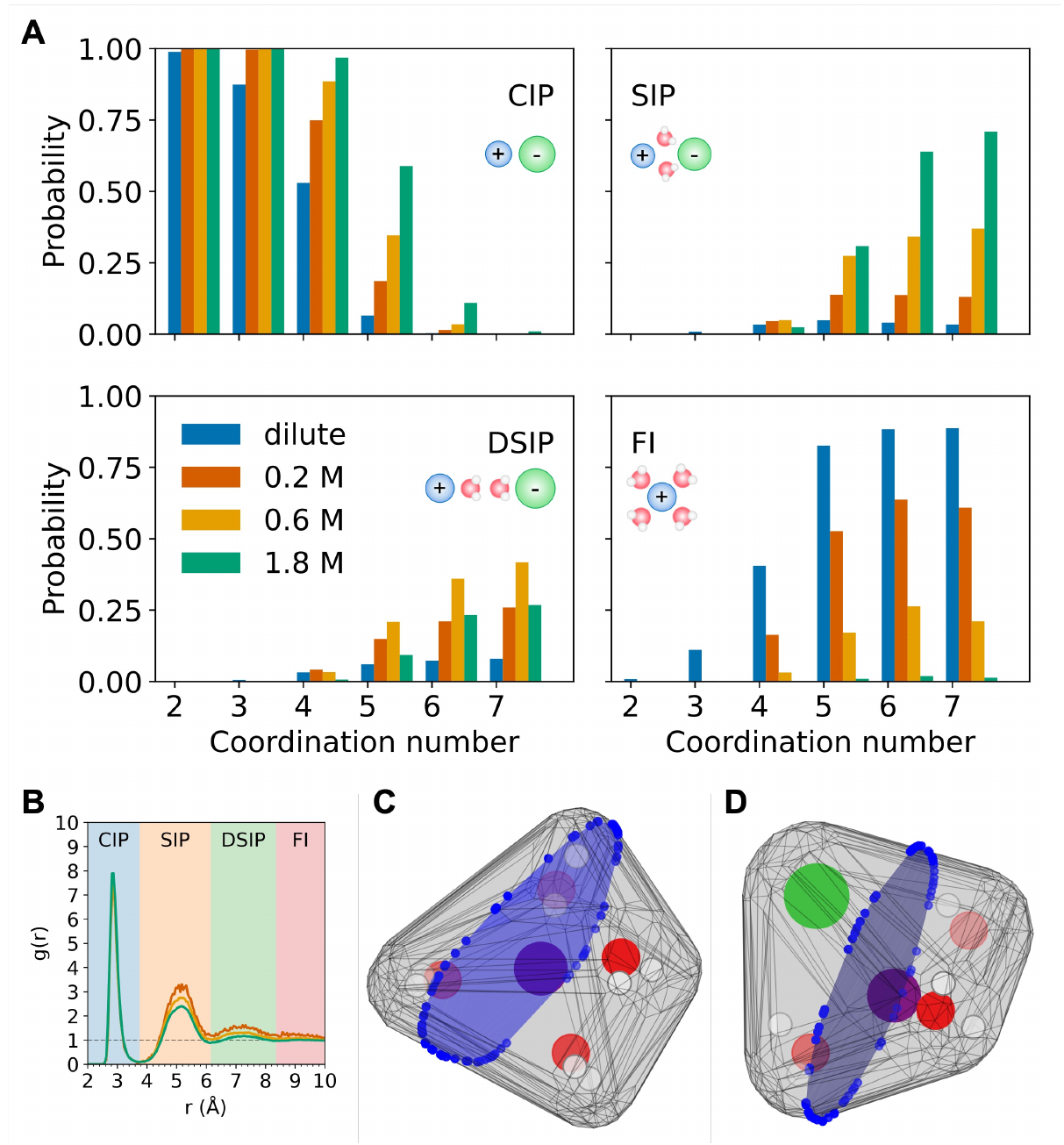}
    \caption{Ion pairing events decrease de-coordination free energies at high concentration. (\textbf{A}) Ion pairing probabilities for Na$^+$ at 1~bar during the umbrella simulations. The frequencies are normalized by the number of frames in a given coordination number state, so the sum over the four ion pairing states for each coordination number is unity. (\textbf{B}) The Na$^+$-Cl$^-$ RDF for 0.2~M (red), 0.6~M (orange), and 1.8~M (green) from unbiased MD are shown with the ion pairing states indicated by colored shading. We observe more CIPs and less SIPs and DSIPs at high concentration. We visualize two representative coordination shells for the biased Na$^+$ ion from umbrella simulations at infinite dilution (\textbf{C}) and 1.8~M (\textbf{D}). These shells were selected from configurations with coordination number 4 and low limiting areas to illustrate where the limiting area lies for shells composed of all water molecules (\textbf{C}) and with an ion pair (\textbf{D}) Na$^+$ is purple. Cl$^-$ is green. Water oxygens are red. Water hydrogens are white. The convex hull formed by the van der Waals spheres of the shell is gray with edges darkened. The limiting cross-sectional area is blue, shown with the points intersecting the convex hull.}
    \label{fig:ion_pairing}
\end{figure}

Our results indicate that high salinity reduces the penalty to de-coordination of cations in solution. To further elucidate the reasons for this reduction in free energy upon de-coordination, we quantify ion-ion interactions with increasing concentration. We classify each configuration as a contact ion pair (CIP), solvent-separated ion pair (SIP), doubly solvent-separated ion pair (DSIP), or free ion (FI), according to the nomenclature presented by Rinne et al.~\cite{rinne_dissecting_2014}. We define these ion pairing events by the distance to the nearest oppositely charged ion. We used the minima in the ion-ion RDF to determine the distance cutoffs as shown in Figure~\ref{fig:ion_pairing}B. Ion-ion distances less than the first minimum are CIPs, between the first and second minima are SIPs, between the second and third minima are DSIPs, and beyond the third minima are FIs. 

We observe significant ion pairing at high concentrations, and we hypothesize that it is these ion pairing events that primarily contribute to the decreased de-coordination free energies calculated from the umbrella sampling simulations. Previous work has also reported increased ion pairing at high concentrations~\cite{rinne_dissecting_2014}. Figure~\ref{fig:ion_pairing}A shows the distribution of ion pairing events for the discrete coordination number states for Na$^+$ at increasing concentration. At low coordination number, Na$^+$ is most likely in a contact ion pair with Cl$^-$ for all concentrations. However, high concentration changes the most likely states at intermediate to high coordination number. Increasing salinity causes the most probable ion pairing state to switch from FI to SIP for coordination number 6, the minimum-energy coordination number for Na$^+$. For a single de-coordination (i.e.~coordination number 5 for Na$^+$), the most probable ion pairing state switches from FI at infinite dilution to CIP at 1.8~M. At high concentration, therefore, the presence of nearby ions decreases the penalty to de-coordinate by directly filling the coordination shell with counterions rather than water molecules. 

We find that cation de-coordination energies decrease with increasing concentration, but anion de-coordination energies do not change. We hypothesize that ion pairing frequency contributes to this difference as well. We show the ion pairing distributions at each discrete coordination number state for Cl$^-$ in Figure~\ref{fig:SI_Cl_ion_pairing}. The ion pairing states are much less sensitive to the coordination number for Cl$^-$ than for Na$^+$. For example, at infinite dilution Cl$^-$ goes from 90.0\% FI at its minimum-energy CN (CN = 7) to 90.6\% FI after 3 de-coordinations (CN = 4), while Na$^+$ goes from 88.3\% FI at its minimum-energy CN (CN = 6) to 11.1\% after 3 de-coordinations (CN = 3). 

We show representative coordination shells with their polyhedrons for Na$^+$ at infinite dilution (Figure~\ref{fig:ion_pairing}C) and 1.8~M (Figure~\ref{fig:ion_pairing}D). Each visualization has coordination number 4, but the high concentration shell is in a CIP. For this configuration (Figure~\ref{fig:ion_pairing}C) at low concentration, the volume of the polyhedron formed by atoms in the shell is 143.7~\AA$^3$, and the limiting area is 25.93~\AA$^2$. The average volume for CN = 4 at infinite dilution is 148~$\pm$~8~\AA$^3$, and the average limiting area is 30~$\pm$~3~\AA$^3$. For this configuration (Figure~\ref{fig:ion_pairing}D) at high concentration, the volume is 136.7~\AA$^3$, and the limiting area is 27.25~\AA$^2$. The average volume for CN = 4 at 1.8~M is 166~$\pm$~13~\AA$^3$, and the average limiting area is 34~$\pm$~3~\AA$^3$. The free energy to achieve the high concentration shell is lower than the low concentration shell. The CIP compensates for the lower coordination number. However, the low concentration shell would have a slightly smaller steric penalty to move through a membrane constriction due to its lower limiting area. Therefore, the decreased de-coordination penalty induced at high salinity would be partially compensated by a higher steric penalty for this configuration. On average, the shell volume for CN = 4 is higher at high concentration than at low concentration, but the restrictive area does not change much. The difference in de-coordination free energy at high concentration could also be due to overall disorder in the shell, but the steric penalty to move through a constriction may not be affected. 

\subsection{Implications for membrane transport}

We argue that ions are unlikely to shed more than two or three molecules from their coordination shells. More de-coordinations will require free energy penalties too high for transport in RO or NF membranes, even when compensated by some interactions with the membrane. Most barriers reported for permeation through these membranes are below 35~kJ/mol~\cite{shefer_applying_2022}, which most closely resemble the bottleneck barriers in the membrane~\cite{schwindt_interpreting_2024}. Therefore, ions will typically follow free energy pathways with barriers below 35~kJ/mol, and full de-coordination would require overcoming significantly higher barriers. 

If an ion were able to de-coordinate more completely, it would require significant compensating interactions with the membrane. However, preliminary evidence suggests that it is unlikely such membrane interactions can compensate for more half of the missing energy. These precise interactions are likely to be rare in highly heterogeneous RO and NF membranes, making it difficult to leverage full de-coordination. One way to approximate these compensating interactions is comparing the de-coordination energy and the dehydration energy, where counterions are allowed to fill the missing hydration shell. Thus, the interactions with counterions act as an upper bound for the compensating membrane interactions. We compare these free energies at high concentration, since many counterions could then compensate for the reduced number of coordinating water molecules. 

We show the resulting free energies using the water-only coordination number for Na$^+$ in Figure~\ref{fig:SI_coordination_definition}B and C. At 1.8~M, we find that the compensating interaction strength to reach coordination numbers 5, 4, and 3 are 0.57, 5.3, and 13.2~kJ/mol respectively, calculated by subtracting the water-only free energy from the all-molecule free energy. These differences are only 53\% for coordination number 5, 64\% for coordination number 4, and 64\% of the barrier for coordination number 3, indicating that even counterions with maximally complementary can only partly compensate for dehydration. Polymer membranes, even charged membranes, will likely have a difficult time providing compensating energetic interactions. 

We showed that high operating pressures do not significantly change the de-coordination free energies nor the shell size. While we did see a trend in the anion de-coordination free energies, the differences are on the 0.1-1 kJ/mol scale, which is small compared to the trends observed for ion size and concentration, or even thermal fluctuations. Additionally, other mechanisms likely show more dominant effects at high pressure. For example, it is more likely that high pressure will influence the de-coordination mechanism via membrane compaction~\cite{lim_demystifying_2024,fan_theory_2024}, rather than through the stability of the coordination shell. Voids in the membrane would shrink, and thus ions would need to achieve even smaller coordination numbers to traverse constrictions. We do not apply a pressure gradient in our simulations, but these effects are small in comparison to the total de-coordination energy. 

Ion de-coordination theoretically can be exploited to improve selectivity of ions of similar size and valency, as different ions show different barriers. In particular, if membranes can be designed to leverage near complete de-coordination, the free energy differences are large enough to significantly affect ion-ion selectivity. Notably, even the first and second de-coordinations differ enough to achieve some level of selective separation. Additionally, for specific ions, elevated pressure and salinity can exaggerate the differences in de-coordination free energies, potentially improving their selectivity. For example, the de-coordination-governed selectivity for Li$^+$/Mg$^{2+}$ could be represented as proportional to the ratio between the barriers for their first two de-coordination, since the barrier to completely de-coordinate would be too high (Equation~\ref{eq:Li_Mg_selectivity}). This ratio is 1.68 at infinite dilution and 2.71 at 1.8~M (values found in Figure~\ref{fig:SI_mono_cation_fe_concentration}A and \ref{fig:SI_all_fe_concentration}A). Thus increased concentration could improve selectivity since the free energy to de-coordinate does not change as much for Mg$^{2+}$ as it does for Li$^+$.

\begin{equation}
    S \propto \frac{\Delta G_{\text{Mg}^{2+}} \left( \text{CN}=4 \right)}{\Delta G_{\text{Li}^+} \left( \text{CN}=2 \right)}
    \label{eq:Li_Mg_selectivity}
\end{equation}

Ion-ion selectivity from de-coordination that is present at lower pressures will be maintained, or even enhanced, at elevated pressure. If membrane compaction decreases the size of constrictions in the polymer, ions will be forced into lower coordination number states, where differences in ion de-coordination free energies are more pronounced. These larger free energy differences would further improve de-coordination-driven selectivity. 

At high salinity, it is easier to reduce the coordination number for cations than at low salinity (Figure~\ref{fig:free_energy_concentration}A). This trend is consistent with experimental reports of increased salt permeability at high salinity, which would correspond to lower barriers for salt transport through the membrane~\cite{davenport_high-pressure_2018,pataroque_salt_2024,chen_transport_2020}. High concentration increases charge screening in aqueous electrolyte solutions, which would destabilize the hydration shell and may contribute to the lower de-coordination free energy for cations~\cite{smith_electrostatic_2016}. 


However, the decreasing ion de-coordination free energy is only significant if there is increased salinity in the local environment. While this will be true in solution or near the solution-membrane interface, it is not clear if there is increased salinity inside the voids of a membrane that is likely less polar than water, even if the external salinity is high. Therefore, the effects of ion salinity ion de-coordination are more relevant for NF membranes with relatively large voids than for RO membranes.

Finally, we characterized the largest cross-sectional area of the shell along its principal axis, which we argue is a better descriptor of the necessary size to traverse nanoscale constrictions than the coordination number. De-coordination typically reduces the sterically-restrictive area, but a lower coordination number does not necessarily require a lower cross-sectional area. Therefore, the barrier to de-coordinate will not necessarily correspond to the barrier to move through a constriction. We thus calculated the free energy as a function of the limiting cross-sectional area, and we show that a moderate reduction in coordination number is easier than a moderate reduction in cross-sectional area. 

\section{Conclusions}

We defined the all-molecule, or total, coordination number in order to probe the barrier to reducing the total transporting volume of a hydrated ion and examined this free energy of de-coordination throughout this paper. As a key enabling technology, we developed a procedure to calculate this barrier in solution with umbrella sampling simulations. We estimated free energies in a discrete collective variable (coordination number) from enhanced sampling simulations biased in a continuous collective variable (coordination number using a switching function). While the continuous coordination number allows for intermediate states between physical states, it introduces ambiguity in the interpretation of the free energies. Molecules near the cutoff contribute partial values to the coordination number. We presented a procedure to reweight to the discrete coordination numbers which is independent of the continuous surrogate for coordination number, as long as that surrogate sufficiently samples the relevant discrete coordination numbers. By reweighting to the discrete distribution, the free energies are more clearly the barriers to transition from the minimum-energy coordination number to lower coordination number states. 

As we discussed earlier in this study, the barriers we report here are higher than those experienced by ions in the membrane due to compensating effects caused by coordination with membrane groups. However, trends at relevant conditions and relative free energies can provide insight into the nanoscale mechanisms governing membrane transport. For example, well-established trends like Na$^+$-K$^+$ selectivity in biological channels and anion permeation barriers are consistent with the trends presented here. Small differences in the de-coordination free energies for Na$^+$ and K$^+$ provide selectivity that can be exploited by precise ion channels~\cite{dudev_factors_2010}. Larger anions show lower de-coordination barriers, which is consistent with experimentally-observed barriers to transmembrane permeation~\cite{epsztein_activation_2019,lu_dehydration-enhanced_2023}. We emphasize that near complete de-coordination will likely require overcoming very high barriers that will not be relevant in amorphous polymer membranes, where fluctuating voids would introduce lower energy pathways~\cite{schwindt_interpreting_2024}. The random, heterogeneous environment would not be precise enough to introduce all the necessary ion-membrane interactions to achieve full de-coordination. We show that even mobile charges (counterions) compensate only partly for the de-coordination. Fixed charges in a polymer membrane would be even less likely to reduce the barrier to complete de-coordination. 

Our results support the conclusions that changes in membrane performance at elevated pressure are not due to changes in ion de-coordination but rather changes in polymer structure from compaction or some other mechanism. We show that high concentration reduces the de-coordination barrier for cations. Cations show significant ion pairing events at high concentration, which contributes to the destabilization of the coordination shell. Nearby ions enable easier de-coordination. However, this trend does not hold for the free energies to reduce the restrictive cross-sectional area of the coordination shell, which do not change significantly with increasing salinity. 

Our work provides a deeper, molecular-level understanding of ion dehydration and how it relates to ion transport in polymer membranes. The trends we observe serve as a useful reference for determining the molecular mechanisms involved in experimental studies of ion transport in novel brine treatment membranes. Although overall we see only small ion-dependent differences in free energies of de-coordination, there are a few exceptions. For example, the differences in de-coordination free energies for Li$^+$ and Mg$^{2+}$ could guide the design of membranes for Li recovery from salt-lake brines. 

\section*{CRediT authorship contribution statement}
\noindent \textbf{Nathanael S. Schwindt:} Conceptualization, Formal Analysis, Methodology, Software, Investigation, Visualization, Writing - Original Draft, Writing - Review \& Editing \textbf{Razi Epsztein:} Conceptualization, Writing - Review \& Editing, Funding acquisition \textbf{Anthony P. Straub:} Conceptualization, Writing - Review \& Editing, Funding acquisition \textbf{Shuwen Yue:} Writing - Review \& Editing \textbf{Michael R. Shirts:} Conceptualization, Resources, Writing - Review \& Editing, Supervision, Funding acquisition

\section*{Declaration of Competing Interest}
\noindent The authors declare that they have no known competing financial interests or personal relationships that could have appeared to influence the work reported in this paper.



\section*{Acknowledgements}
This material is based upon work supported by the National Science Foundation under Grant No. CBET-2136835 and the United States-Israel Binational Science Foundation (BSF), Jerusalem, Israel (grant No. 2021615). This work used Bridges-2 at Pittsburgh Supercomputing Center through allocation chm230020p from the Advanced Cyberinfrastructure Coordination Ecosystem: Services \& Support (ACCESS) program, which is supported by National Science Foundation grants \texttt{\#}2138259, \texttt{\#}2138286, \texttt{\#}2138307, \texttt{\#}2137603, and \texttt{\#}2138296.

\appendix
\section{Supplementary Materials}

\bibliographystyle{elsarticle-num}
\bibliography{dehydration}

\pagebreak

\captionsetup[figure]{labelfont={bf}, labelsep=period, name={Fig.}}
\captionsetup[table]{labelfont={bf}, labelsep=period, name={Table}}
\renewcommand\thesection{S\arabic{section}}
\renewcommand\thefigure{S\arabic{figure}}
\renewcommand\thetable{S\arabic{table}}
\renewcommand\theequation{S\arabic{equation}}

\setcounter{figure}{0}
\setcounter{equation}{0}
\setcounter{page}{1}
\setcounter{section}{0}

\section*{\Large Supplementary Materials}

\section{All-molecule coordination number better represents the true size of the coordination shell} \label{s:SI_dehydration_definition}

 \begin{figure*}[ht!]
    \centering
    \includegraphics[width=\textwidth]{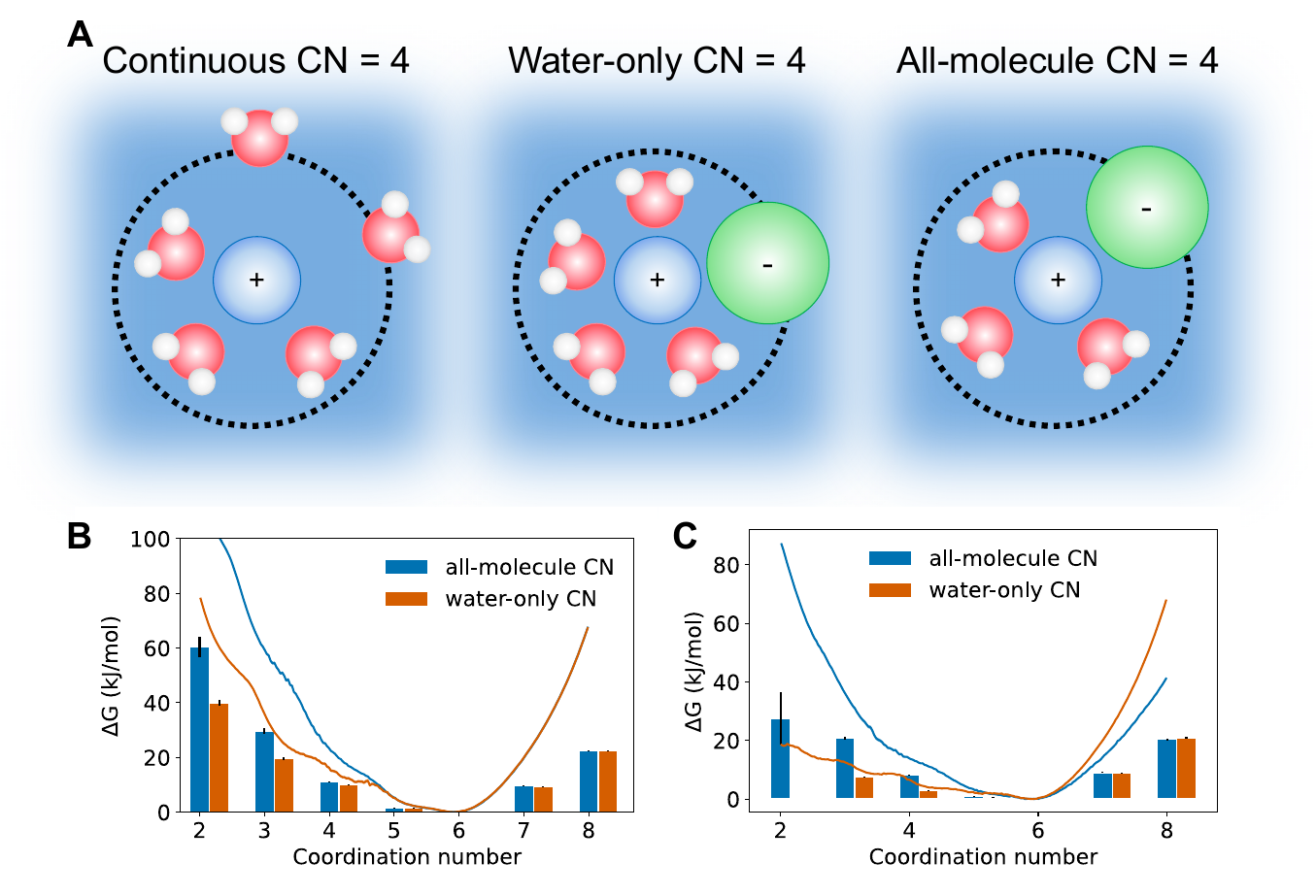}
    \caption{Defining de-coordination using the all-molecule coordination number (CN) isolates the free energy to reduce the size of the coordination shell. (\textbf{A}) Visualizations of the different CN definitions show cations, anions, and water molecules in relation to the shell cutoff (black dashed line). Continuous CN allows for partial contributions from molecules near the shell cutoff. Water-only CN, or hydration number, allows counterions to enter the shell without changing the CN. All-molecule CN creates voids in the shell upon de-coordination. (\textbf{B}) Continuous CN and discrete CN free energies at infinite dilution show a lower penalty to de-coordination when the water-only CN is used. (\textbf{C}) High concentration (1.8~M) further lowers the de-coordination free energy since there are many nearby ions that can fill the shell.}
    \label{fig:SI_coordination_definition}
\end{figure*}

We explored definitions of the coordination number in order to more clearly define the mechanism important for membrane transport and the associated free energy barriers we estimated. We show theoretical coordination shells with coordination number 4 for these definitions in Figure~\ref{fig:SI_coordination_definition}. 

We began with the hydration shell, or only counting coordinating water molecules. When biasing the simulation in the hydration number, counterions can enter the hydration shell to fill the void left by the dehydrated water molecule as shown in the middle panel of Figure~\ref{fig:SI_coordination_definition}A. As a result, the free energy to reduce the hydration number is low, especially at high concentration. We then defined the coordination number as the total number of molecules within an ion's shell, including any type of ion. As a result, reducing the coordination number always corresponded to fewer molecules in the shell. 

We argue that reducing the number of molecules that must transport with a given ion is more physically relevant than only considering the water molecules. We also illustrate coordination number 4 calculated by the switching function in order to have a continuous collective variable that is a function of the atomic coordinates, which is necessary to apply bias with Plumed. We provide more information on this definition in Supplementary Information Section~\ref{s:SI_switching_function}. We calculated the continuous free energy surface and the free energies of coordination number states in the same way for both definitions of coordination number, and we provide an example of these calculations on Github at \url{https://github.com/shirtsgroup/solvation_shells/}.

We also calculated the free energy surface using the water-only coordination number with an additional restraint applied to the counterion coordination number, i.e. the number of counterions within the hydration shell radius. The restraint in counterion coordination prevents the simulation from sampling configurations with a counterion in the hydration shell. 

\begin{figure*}[ht!]
    \centering
    \includegraphics[width=\textwidth]{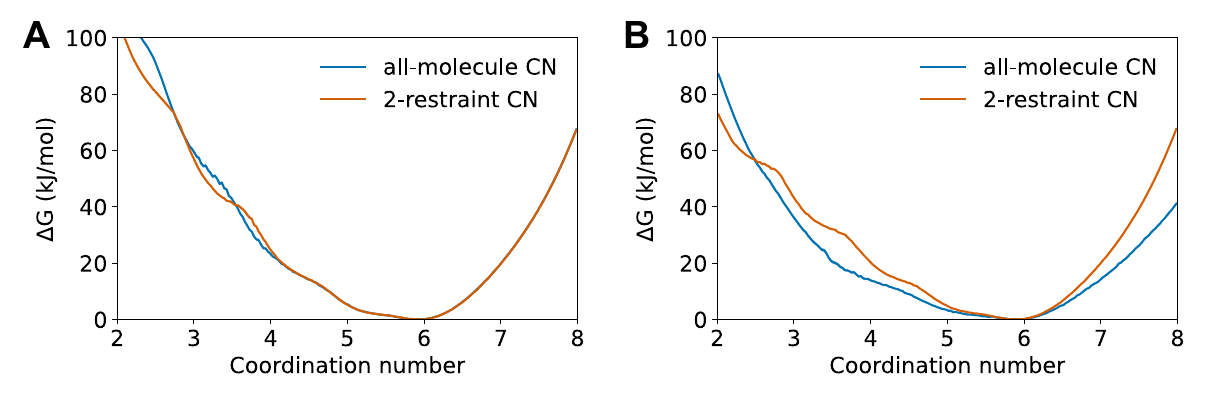}
    \caption{Adding a second restraint to the counterion coordination number significantly increases the free energy to de-coordinate at high concentration. (\textbf{A}) We compare the free energy surface in the continuous, water-only coordination number at infinite dilution and 1~bar for the all-molecule coordination number defintion and the simulations with 2 restraints. The free energies are identical except at very low coordination number. (\textbf{B}) The free energies for the 2-restraint coordination number at 1.8~M and 1~bar are much higher than those for the all-molecule coordination number since ion pairing is more favorable at low coordination number. However, the second restraint prevents sampling ion pairing events.}
    \label{fig:SI_2_restraints}
\end{figure*}

Figure~\ref{fig:SI_2_restraints} shows the continuous free energy surface for the all-molecule coordination number definition compared to that calculated with the second restraint at low (\textbf{A}) and high (\textbf{B}) concentration. The free energies are very similar at low concentration, since not many configurations can have counterions in the shell. The free energy calculated with the second restraint is significantly higher at high concentration, since it is more favorable to fill the shell with counterions at low coordination number. To calculate the free energy with the second restraint, we reweight the simulations accounting for both harmonic restraints, but calculate the free energy surface in a single collective variable -- the water-only coordination number. We must have sampling overlap between the water-only coordination number umbrellas, but we include only one umbrella in counterion coordination.

\clearpage
\pagebreak

\section{Sensitivity to switching function} \label{s:SI_switching_function}

 \begin{figure*}[ht!]
    \centering
    \includegraphics[width=0.8\textwidth]{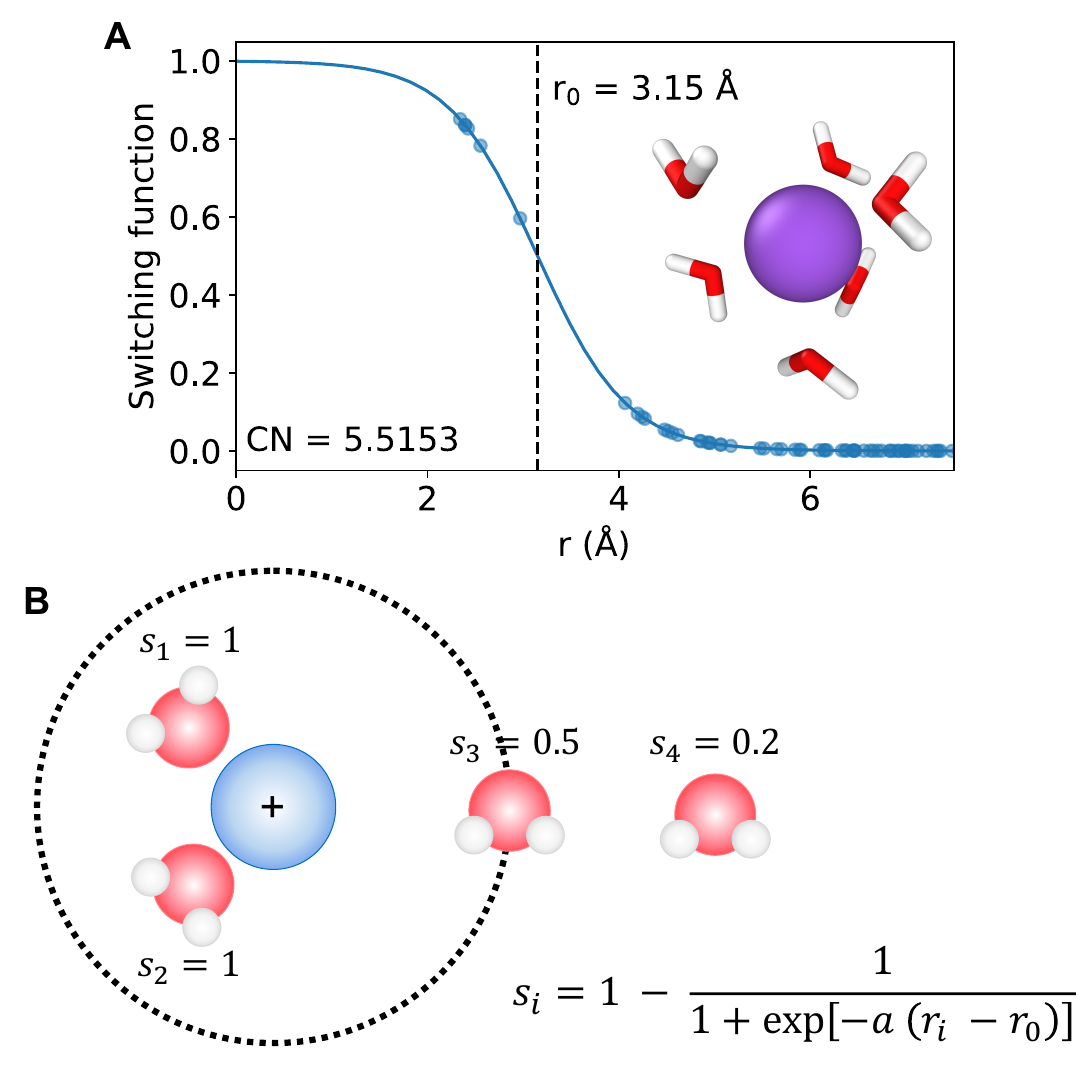}
    \caption{Switching function includes partial contributions from molecules near the cutoff. (\textbf{A}) We show the switching function applied to a single frame from the umbrella simulations of Na$^+$ at infinite dilution and 1~bar using $a$ = 21.5. The inlaid image is the coordination shell of the sodium ion. Notably, the shell clearly contains six water molecules, while the continuous coordination number using the switching function is 5.5153. (\textbf{B}) We show a conceptualization of how molecules can partially contribute to the continuous coordination number. We include the switching function equation. The continuous coordination number is the sum of all $s_i$ over all molecules in the system.}
    \label{fig:SI_switching_function}
\end{figure*}

The switching function parameter $a$ (equation in Figure~\ref{fig:SI_switching_function}B) is a nonphysical parameter that controls the steepness of the switching function, and it is necessary for numerical stability when applying bias to the coordination number in Plumed. Plumed requires the collective variable to be a continuous function of the atomic coordinates at a given time step. Since the $a$ is not physically-motivated, we tested the sensitivity of the free energy calculations to this parameter. 

\begin{figure*}[ht!]
    \centering
    \includegraphics[width=\textwidth]{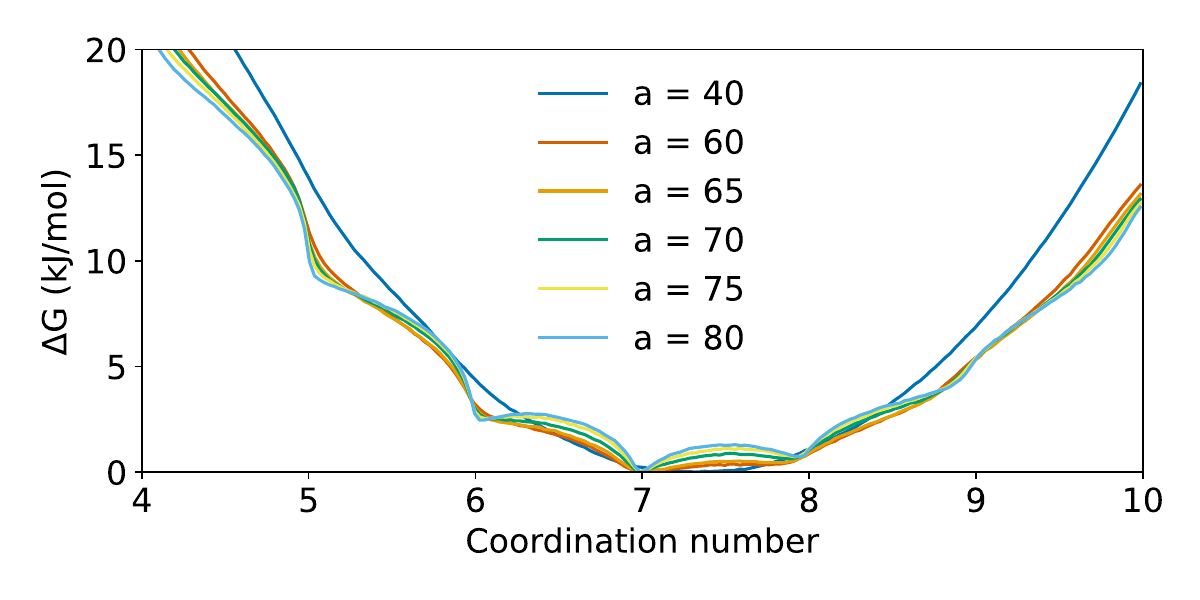}
    \caption{High switching function parameters ($a$) create sharp minima near integer values of coordination number, and low $a$ shifts the free energy minima to higher coordination number. Here, we show the continuous coordination number free energy surfaces for many switching functions for Rb$^+$ at 1~bar and infinite dilution. These free energy surfaces were calculated from short (10~ns) umbrella sampling simulations using the configuration shown in Section~\ref{s:SI_umbrella_configs} for Rb$^+$.}
    \label{fig:SI_fes_a_values}
\end{figure*}

Figure~\ref{fig:SI_switching_function}A shows the switching function plotted for single frame from the umbrella simulations of Na$^+$. Physically, the coordination number is 6 as shown in the inlaid image, but the switching function calculated coordination number is 5.5153. Each molecule's contribution to the coordination number is shown as a scatter point, and points near the cutoff only contribute $\sim$0.5 as shown in the illustration in Figure~\ref{fig:SI_switching_function}B. Increasing $a$ will make the switching function closer to a discrete counting function, but if it is too steep, it causes numerical instability. High $a$ values also result in sharp minima in the free energy surface as shown in Figure~\ref{fig:SI_fes_a_values}. On the other hand, decreasing $a$ too much moves the free energy minimum farther from the true integer value, which can be seen in the $a$ = 40 case on Figure~\ref{fig:SI_fes_a_values}.

\begin{figure*}[ht!]
    \centering
    \includegraphics[width=\textwidth]{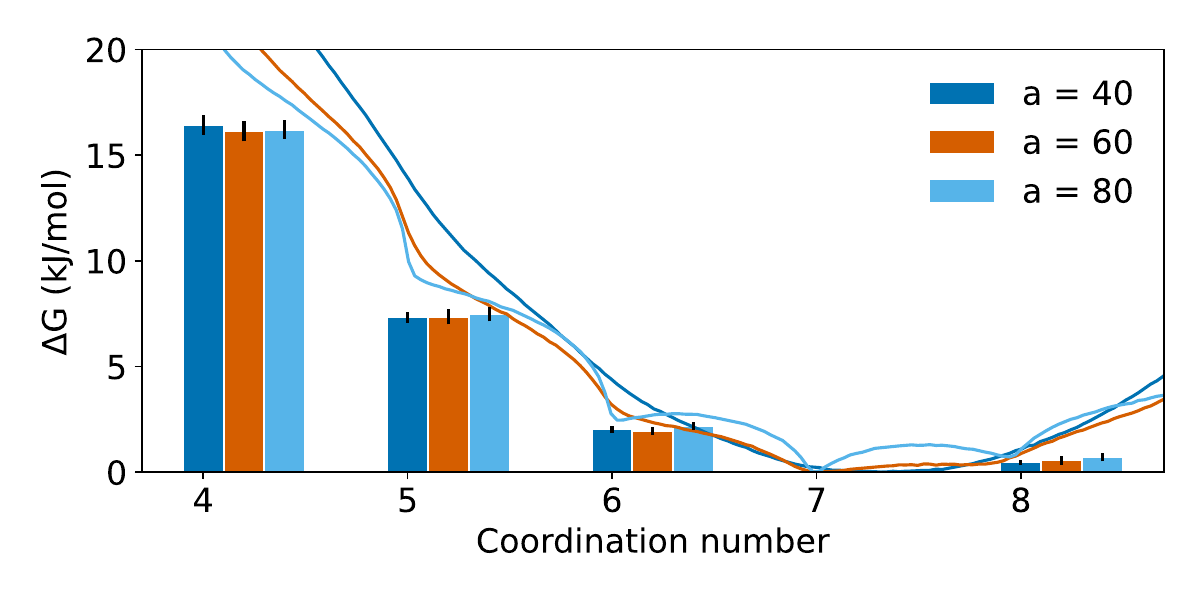}
    \caption{Reweighting to the discrete distribution eliminates the dependence on the switching function. We show the continuous coordination number free energy surfaces and the corresponding discrete free energies for three switching functions for Rb$^+$ at 1~bar and infinite dilution. We chose three switching functions that ranged the parameter space we tested for clarity.}
    \label{fig:SI_discrete_a_dependence}
\end{figure*}

We largely eliminated the dependence on the switching function parameter by reweighting the continuous coordination number configurations to the discrete distribution. Following the procedure described in the main text, we calculated the free energies of the (discrete) coordination number states. These free energies are statistically indistinguishable for all $a$ values. Figure~\ref{fig:SI_discrete_a_dependence} shows the free energies of coordination number states for three switching functions for Rb$^+$. 

\clearpage
\pagebreak

\section{Dependence on force field parameters} \label{s:SI_force_field}

\begin{figure*}[ht!]
    \centering
    \includegraphics[width=\textwidth]{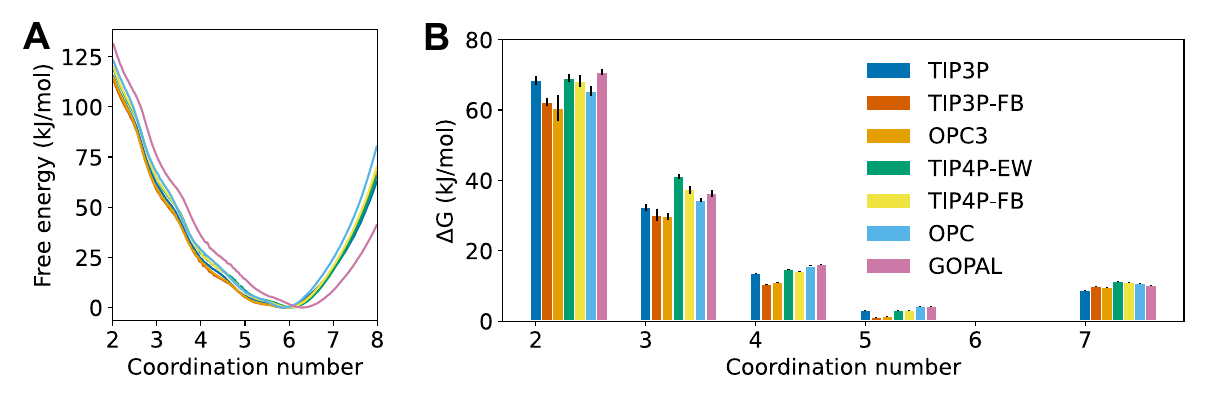}
\caption{Free energy as a function of coordination number is not strongly dependent on the water model. (\textbf{A}) Free energy surfaces in continuous coordination number are qualitatively similar for different waters models, unless there are not optimized ion parameters, as in the case of GOPAL. (\textbf{B}) Free energies for the discrete coordination number states are qualitatively similar. All water models are tested with Na$^+$ at infinite dilution and 1~bar, using optimized ion parameters as discussed in the main text.}
    \label{fig:SI_water_models}
\end{figure*}

Since non-polarizable water models are less accurate with ions, we examined how seven different non-polarizable water models affected the de-coordination free energies for Na$^+$. We tested a range of three- and four-point water models as described in the main text. Figure~\ref{fig:SI_water_models}A shows the free energy surfaces in the continuous coordination number for all water models tested, and Figure~\ref{fig:SI_water_models}B shows the free energies of the coordination number states. Both the free energy surfaces and the discrete free energies are qualitatively similar for all water models. We used OPC3 water with ions tuned for OPC3~\cite{sengupta_parameterization_2021} for all our results. It is consistent with the other water models, and it produces the best pure water properties of the less expensive three-point models. 

\begin{figure}[H]
    \centering
    \includegraphics[width=\textwidth]{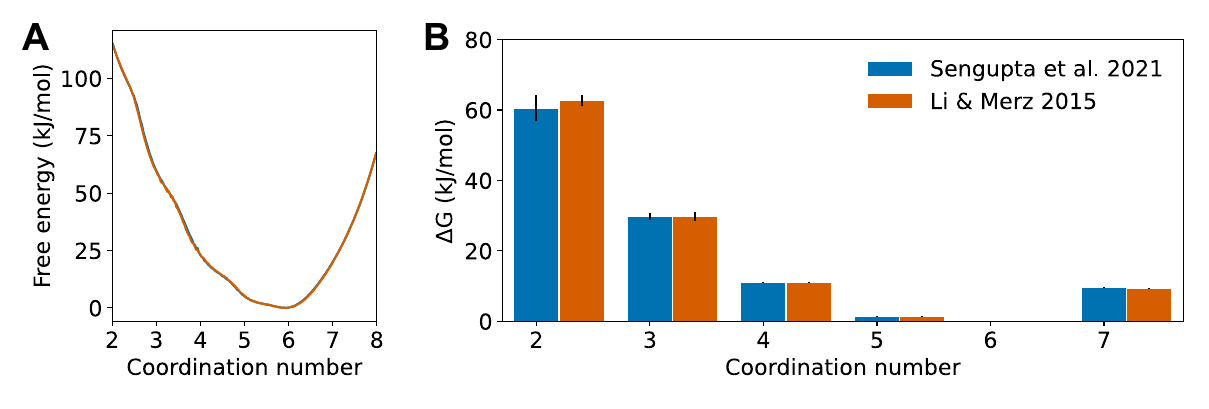}
    \caption{Free energy as a function of coordination number does not change with similarly optimized ion parameters. We calculated the free energies using OPC3 water with ion parameters optimized for OPC3 (Sengupta et al. 2021)~\cite{sengupta_parameterization_2021} and those optimized for TIP3P (Li \& Merz 2015)~\cite{li_systematic_2015}. (\textbf{A}) Free energy surfaces in continuous coordination number are indistinguishable for the different ion parameters. (\textbf{B}) Free energies for the discrete coordination number states are the same within error. Ion parameters are tested with Na$^+$ at infinite dilution and 1~bar.}
    \label{fig:SI_ion_params}
\end{figure}

With our chosen water model, we tested how sensitive the de-coordination free energies are to the ion parameters. We calculated the free energies for Na$^+$ in OPC3 water using ion parameters optimized for OPC3 (Sengupta et al. 2021)~\cite{sengupta_parameterization_2021} and those optimized for TIP3P (Li \& Merz 2015)~\cite{li_systematic_2015}. Figure~\ref{fig:SI_ion_params} depicts the continuous coordination number free energy surfaces (\textbf{A}) and free energies of the discrete coordination number states (\textbf{B}) for the two ion parameter sets. 

\clearpage
\pagebreak

\section{RDFs, limiting area distributions, and de-coordination free energies for all ions as a function of pressure} \label{s:SI_complete_pressure}

 \begin{figure}[H]
    \centering
    \includegraphics[width=0.75\textwidth]{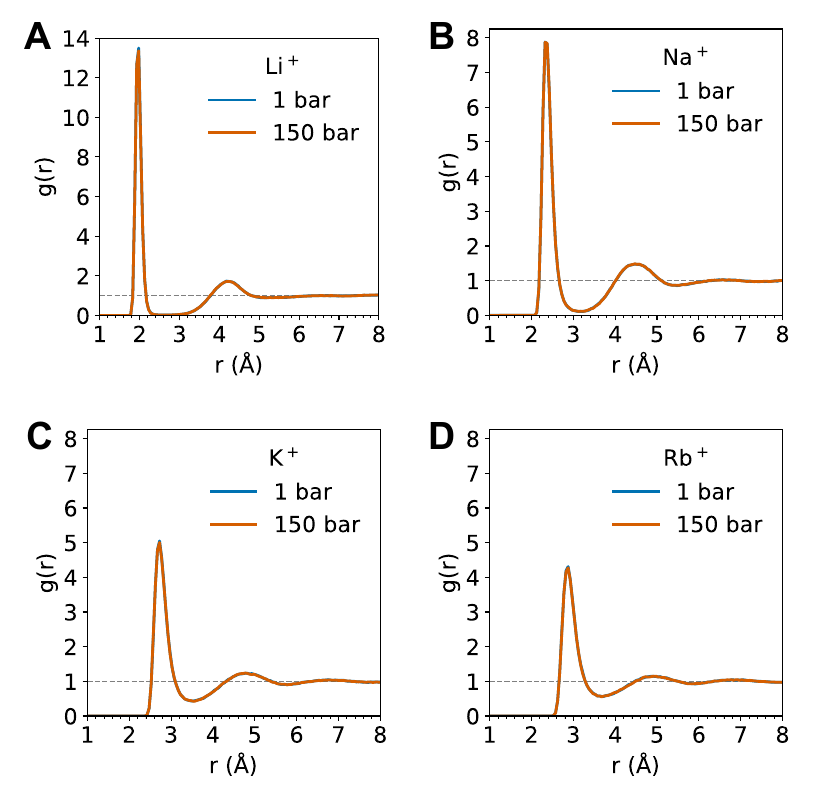}
    \caption{Monovalent cation RDFs at 1~bar and 150~bar do not differ significantly. Here, we compare the ion-water oxygen RDFs for Li$^+$ (\textbf{A}), Na$^+$ (\textbf{B}), K$^+$ (\textbf{C}), and Rb$^+$ (\textbf{D}). Shell structure data were calculated from 20~ns trajectories at infinite dilution.}
    \label{fig:SI_mono_cation_RDFs_pressure}
\end{figure}

 \begin{figure}[H]
    \centering
    \includegraphics[width=\textwidth]{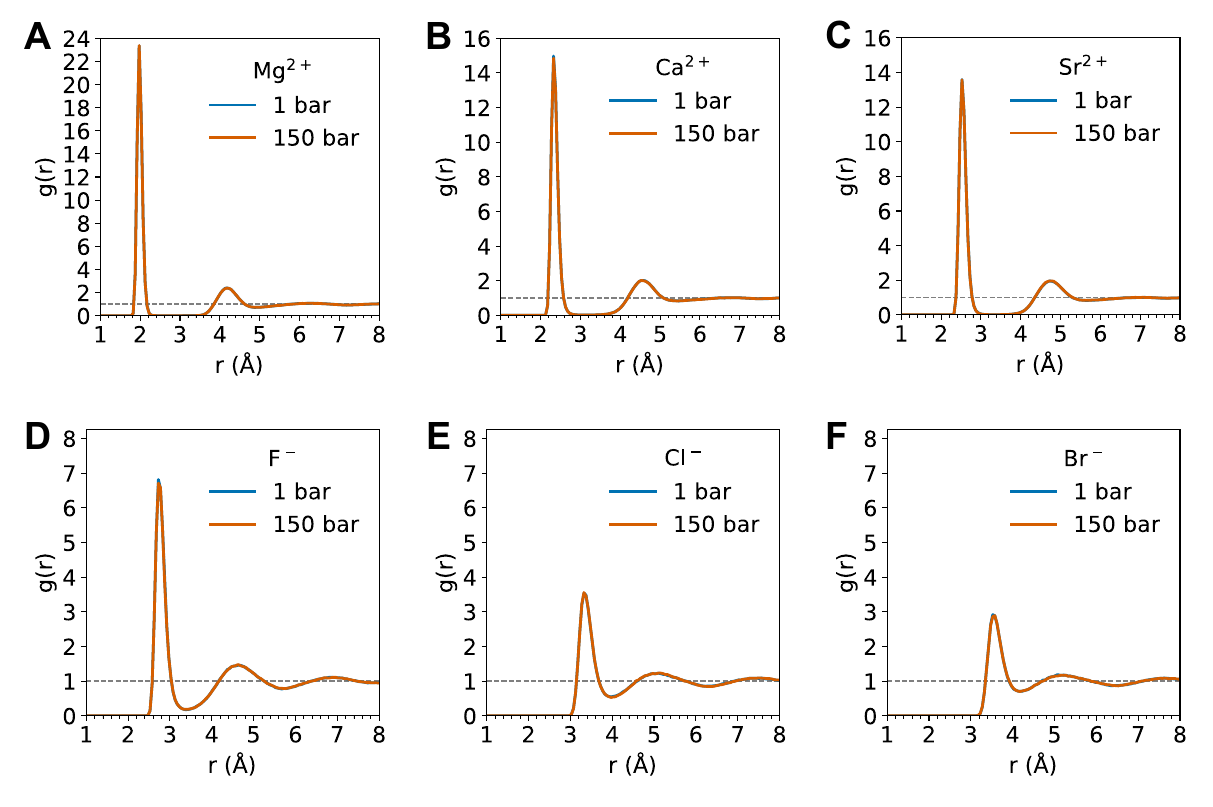}
    \caption{Divalent cation and monovalent anion RDFs at 1~bar and 150~bar do not differ significantly. Here, we compare the ion-water oxygen RDFs for Mg$^{2+}$ (\textbf{A}), Ca$^{2+}$ (\textbf{B}), Sr$^{2+}$ (\textbf{C}), F$^-$ (\textbf{D}), Cl$^-$ (\textbf{E}), and Br$^-$ (\textbf{F}). RDFs were calculated from 20~ns trajectories at infinite dilution.}
    \label{fig:SI_all_RDFs_pressure}
\end{figure}

 \begin{figure}[H]
    \centering
    \includegraphics[width=0.75\textwidth]{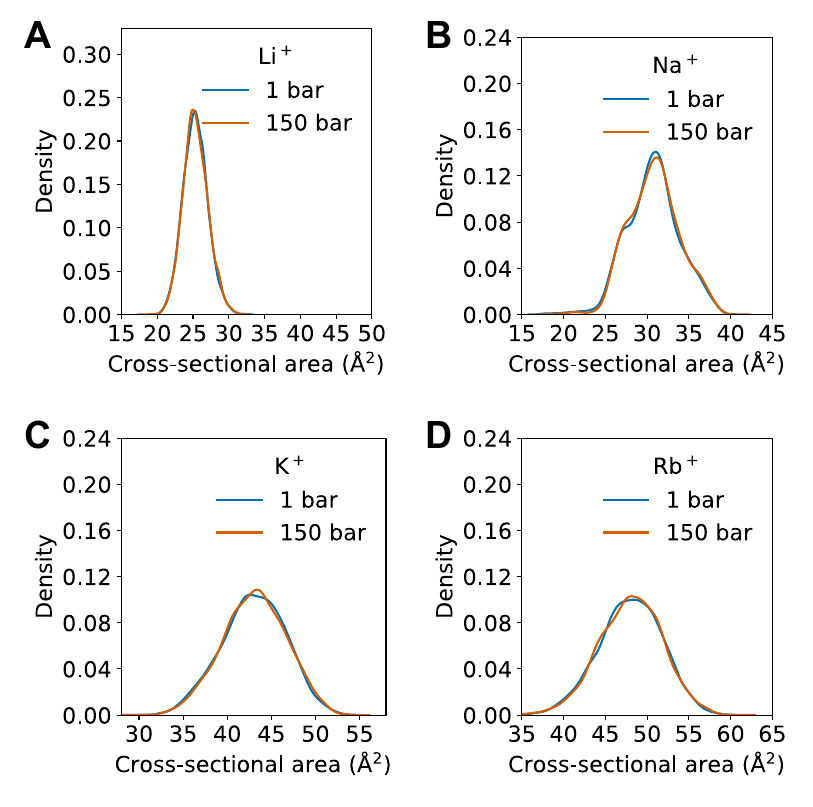}
    \caption{Monovalent cation limiting area distributions at 1~bar and 150~bar do not differ significantly. Smaller cations have narrower distributions. Here, we compare distributions of the maximum cross-sectional area along the principal axis of a polyhedron formed by the atoms in the coordination shell for Li$^+$ (\textbf{A}), Na$^+$ (\textbf{B}), K$^+$ (\textbf{C}), and Rb$^+$ (\textbf{D}). Distributions are shown using kernel density estimations with Scott's Rule to determine the bandwidth~\cite{scott_multivariate_1992}. Cross-sectional areas were calculated from 20~ns trajectories at infinite dilution.}
    \label{fig:SI_mono_cation_areas_pressure}
\end{figure}

 \begin{figure}[H]
    \centering
    \includegraphics[width=\textwidth]{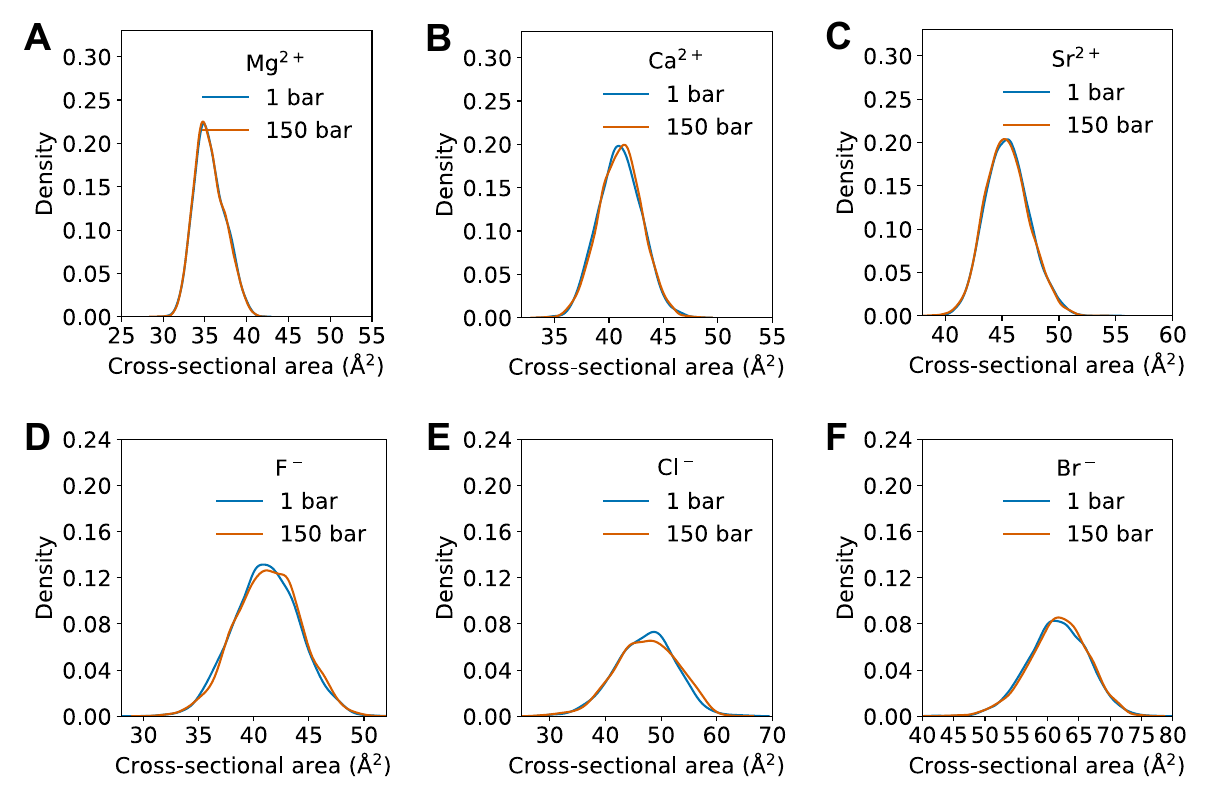}
    \caption{Divalent cation limiting area distributions at 1~bar and 150~bar do not differ significantly. Anion limiting area distributions have slightly more variance at high pressure. Here, we compare distributions of the maximum cross-sectional area along the principal axis of a polyhedron formed by the atoms in the coordination shell for Mg$^{2+}$ (\textbf{A}), Ca$^{2+}$ (\textbf{B}), Sr$^{2+}$ (\textbf{C}), F$^-$ (\textbf{D}), Cl$^-$ (\textbf{E}), and Br$^-$ (\textbf{F}). Distributions are shown using kernel density estimations with Scott's Rule to determine the bandwidth~\cite{scott_multivariate_1992}. Cross-sectional areas were calculated from 20~ns trajectories at infinite dilution.}
    \label{fig:SI_all_areas_pressure}
\end{figure}

 \begin{figure}[H]
    \centering
    \includegraphics[width=0.75\textwidth]{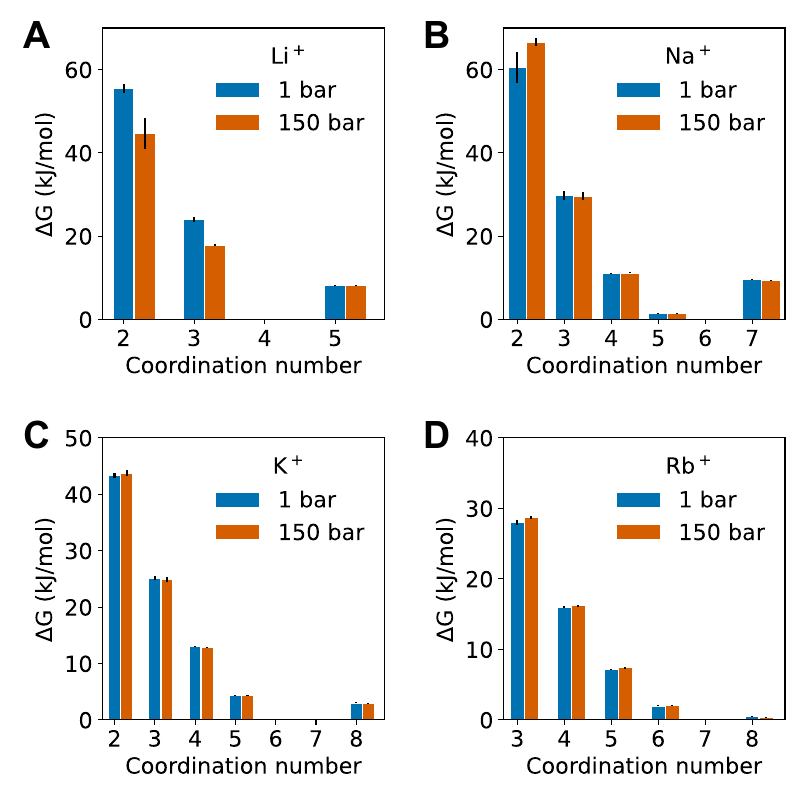}
    \caption{Monovalent cation free energies of coordination number states are not strongly dependent on pressure. De-coordination free energies for Li$^+$ are lower at high pressure. Here, we compare the de-coordination free energies at infinite dilution and increasing pressure -- 1~bar and 150~bar. The monovalent cations we test are Li$^+$ (\textbf{A}), Na$^+$ (\textbf{B}), K$^+$ (\textbf{C}), and Rb$^+$ (\textbf{D}).}
    \label{fig:SI_mono_cation_fe_pressure}
\end{figure}

\begin{figure}[H]
    \centering
    \includegraphics[width=\textwidth]{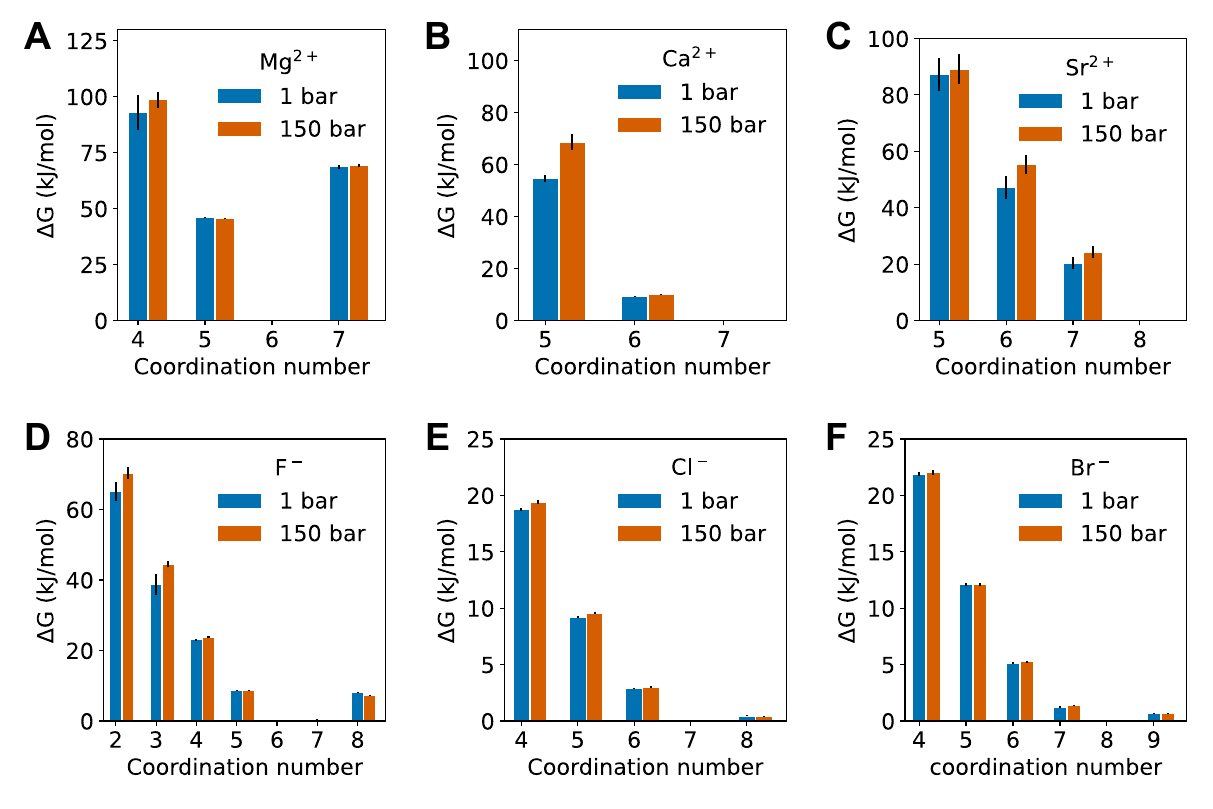}
    \caption{Divalent cation and monovalent anion free energies of coordination number states are slightly higher at 150~bar than at 1~bar. Mg$^{2+}$ free energies do not show a dependence on pressure. Here, we compare the de-coordination free energies at infinite dilution and increasing pressure -- 1~bar and 150~bar. The divalent cations we test are Mg$^{2+}$ (\textbf{A}), Ca$^{2+}$ (\textbf{B}), and Sr$^{2+}$ (\textbf{C}). The monovalent anions we test are F$^-$ (\textbf{D}), Cl$^-$ (\textbf{E}), and Br$^-$ (\textbf{F}).}
    \label{fig:SI_all_fe_pressure}
\end{figure}

\clearpage
\pagebreak

\section{RDFs, limiting area distributions, and de-coordination free energies for all ions as a function of concentration} \label{s:SI_complete_concentration}

\begin{figure}[H]
    \centering
    \includegraphics[width=0.75\textwidth]{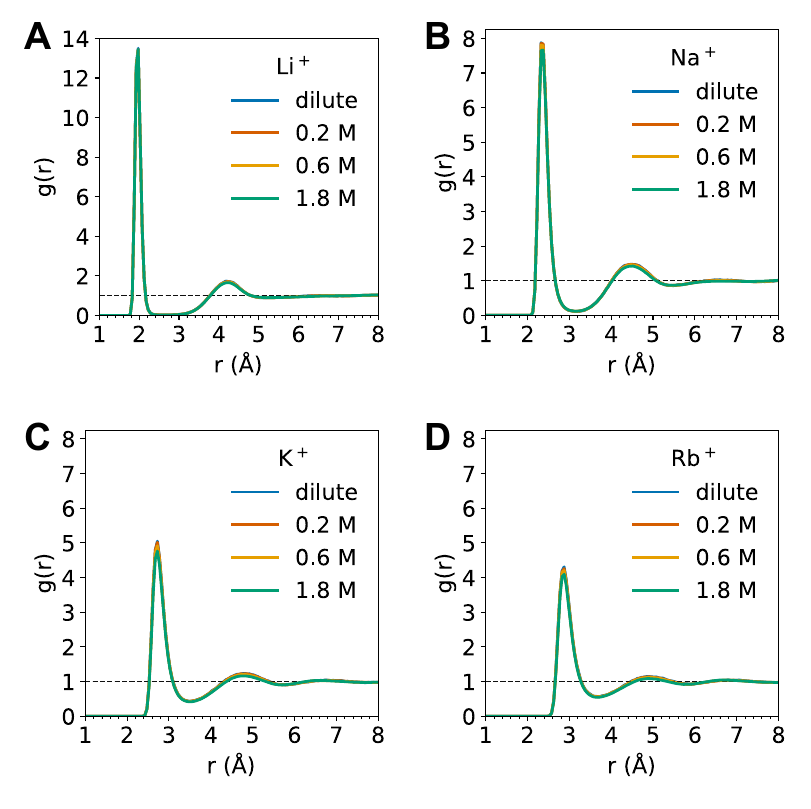}
    \caption{Monovalent cation RDFs at infinite dilution, 0.2~M, 0.6~M, and 1.8~M show increasing density at the ion-oxygen distance with increasing concentration. This peak corresponds to tightly coordinated water molecules. Otherwise, the radial shell structure does not change with concentration. Here, we compare the ion-water oxygen RDFs for Li$^+$ (\textbf{A}), Na$^+$ (\textbf{B}), K$^+$ (\textbf{C}), and Rb$^+$ (\textbf{D}). RDFs were calculated from 20~ns trajectories at 1~bar.}
    \label{fig:SI_mono_cation_RDFs_concentration}
\end{figure}

\begin{figure}[H]
    \centering
    \includegraphics[width=\textwidth]{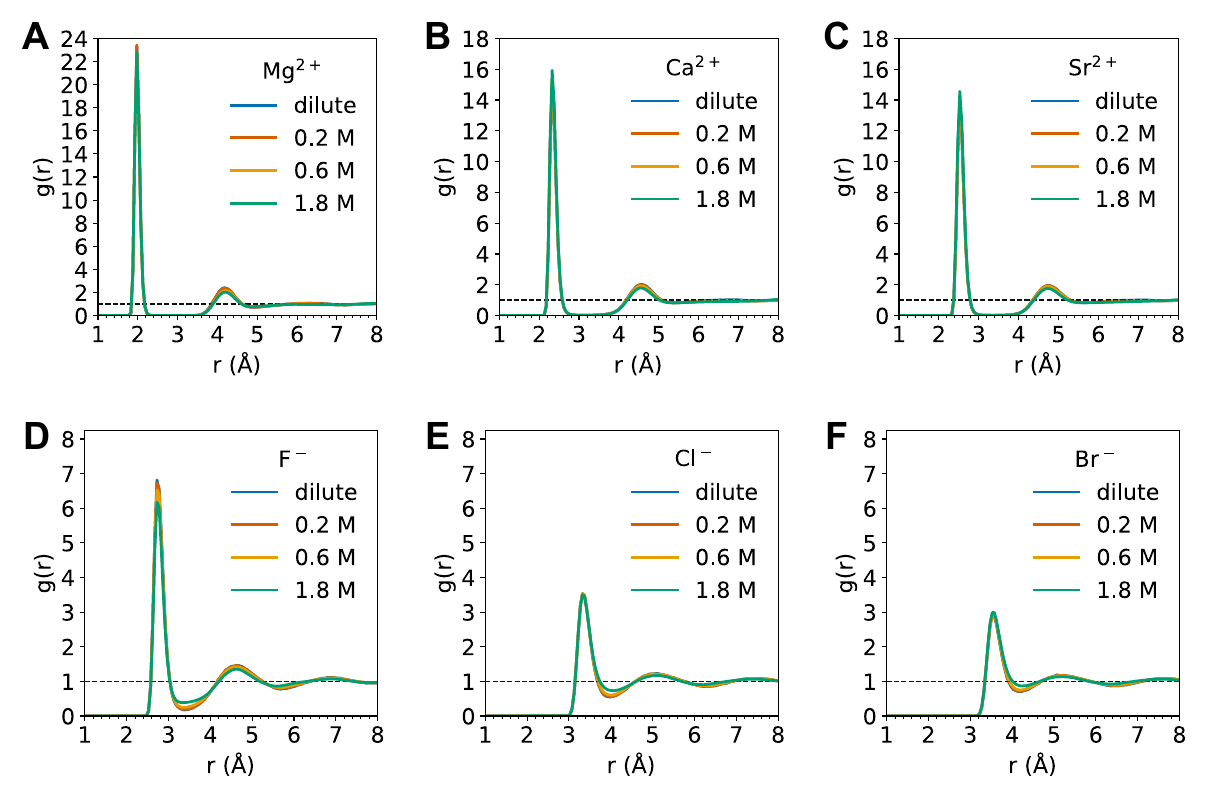}
    \caption{Divalent cation RDFs at infinite dilution, 0.2~M, 0.6~M, and 1.8~M show increasing water density in the second shell, and monovalent anion RDFs show decreasing density near the cutoff with increasing salinity. F$^-$ experiences more tightly coordinated water molecules at higher concentration. Here, we compare the ion-water oxygen RDFs for Mg$^{2+}$ (\textbf{A}), Ca$^{2+}$ (\textbf{B}), Sr$^{2+}$ (\textbf{C}), F$^-$ (\textbf{D}), Cl$^-$ (\textbf{E}), and Br$^-$ (\textbf{F}). RDFs were calculated from 20~ns trajectories at 1~bar.}
    \label{fig:SI_all_RDFs_concentration}
\end{figure}

\begin{figure}[H]
    \centering
    \includegraphics[width=0.75\textwidth]{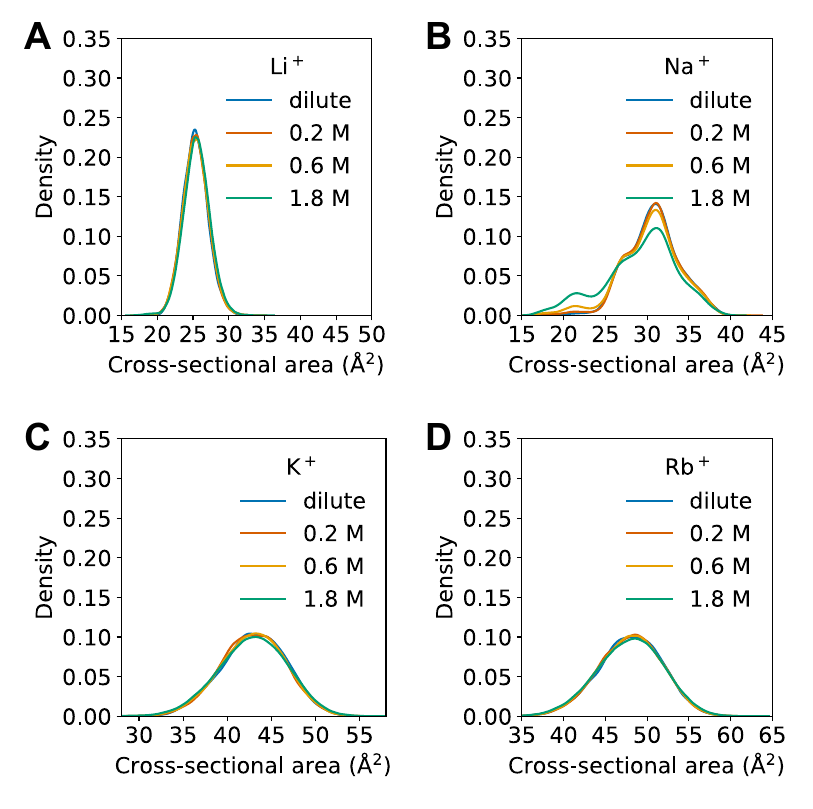}
    \caption{Monovalent cation limiting area distributions at infinite dilution, 0.2~M, 0.6~M, and 1.8~M do not differ significantly, except for Na$^+$, which shows more low-area configurations at high concentration. Here, we compare distributions of the maximum cross-sectional area along the principal axis of a polyhedron formed by the atoms in the coordination shell for Li$^+$ (\textbf{A}), Na$^+$ (\textbf{B}), K$^+$ (\textbf{C}), and Rb$^+$ (\textbf{D}). Distributions are shown using kernel density estimations with Scott's Rule to determine the bandwidth~\cite{scott_multivariate_1992}. Cross-sectional areas were calculated from 20~ns trajectories at infinite dilution.}
    \label{fig:SI_mono_cation_areas_concentration}
\end{figure}

\begin{figure}[H]
    \centering
    \includegraphics[width=\textwidth]{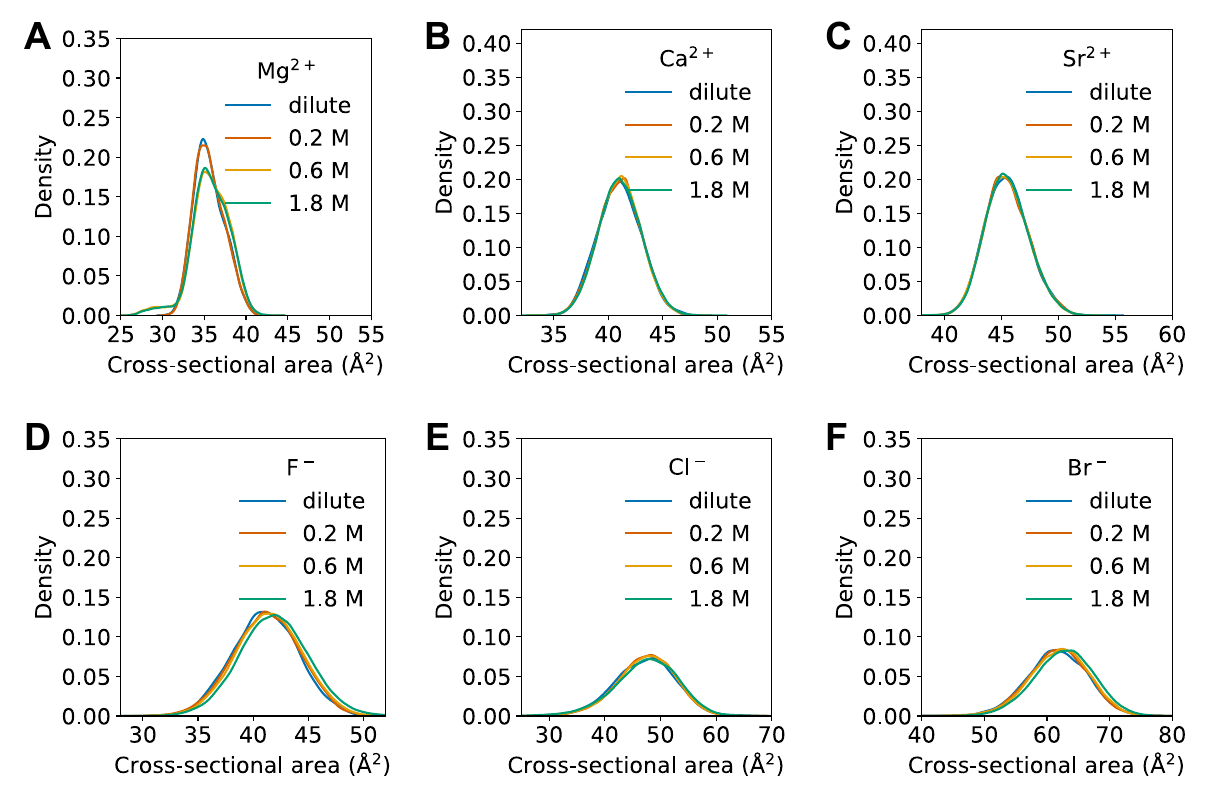}
    \caption{Divalent cation and monovalent anion limiting area distributions at infinite dilution, 0.2~M, 0.6~M, and 1.8~M do not differ significantly, except for Mg$^{2+}$, which shows more low-area configurations at high concentration. Here, we compare distributions of the maximum cross-sectional area along the principal axis of a polyhedron formed by the atoms in the coordination shell for Mg$^{2+}$ (\textbf{A}), Ca$^{2+}$ (\textbf{B}), Sr$^{2+}$ (\textbf{C}), F$^-$ (\textbf{D}), Cl$^-$ (\textbf{E}), and Br$^-$ (\textbf{F}). Distributions are shown using kernel density estimations with Scott's Rule to determine the bandwidth~\cite{scott_multivariate_1992}. Cross-sectional areas were calculated from 20~ns trajectories at infinite dilution.}
    \label{fig:SI_all_areas_concentration}
\end{figure}

\begin{figure}[H]
    \centering
    \includegraphics[width=0.75\textwidth]{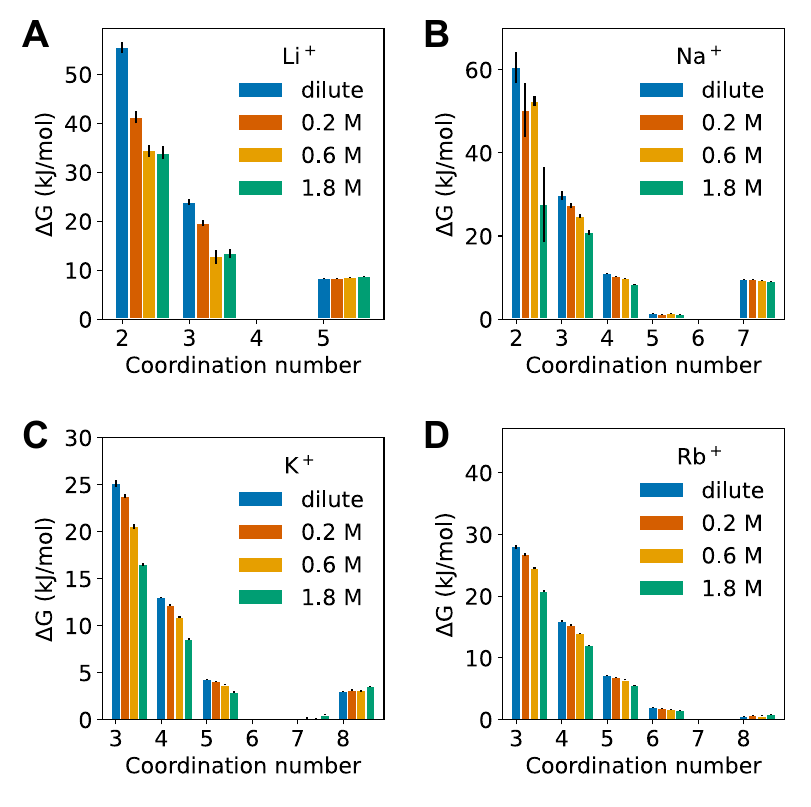}
    \caption{Monovalent cation free energies of coordination number states at infinite dilution, 0.2~M, 0.6~M, and 1.8~M decrease with increasing concentration. Here, we compare the de-coordination free energies at 1~bar for Li$^+$ (\textbf{A}), Na$^+$ (\textbf{B}), K$^+$ (\textbf{C}), and Rb$^+$ (\textbf{D}).}
    \label{fig:SI_mono_cation_fe_concentration}
\end{figure}

\begin{figure}[H]
    \centering
    \includegraphics[width=\textwidth]{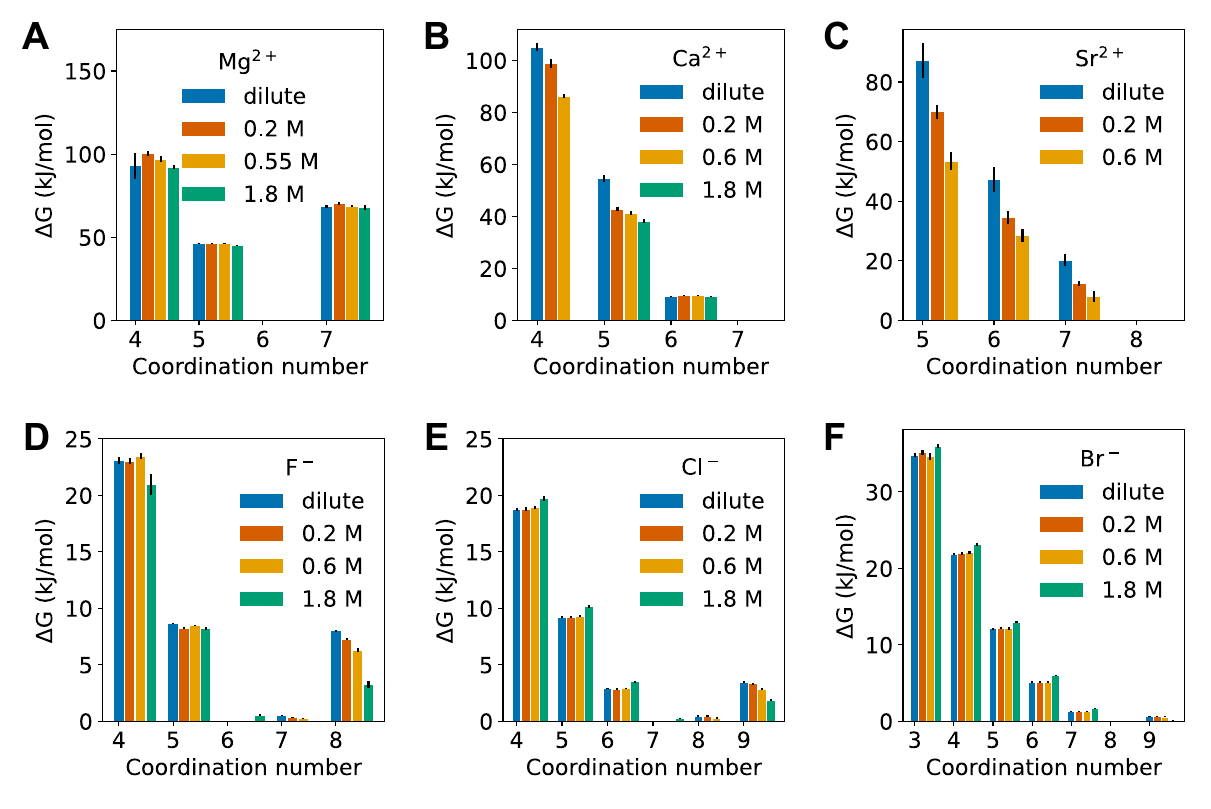}
    \caption{Divalent cation free energies of coordination number states at infinite dilution, 0.2~M, 0.6~M, and 1.8~M decrease with increasing concentration, and monovalent anion free energies are unchanged. The minimum-energy coordination number increases for anions at 1.8~M. Here, we compare the de-coordination free energies at 1~bar for Mg$^{2+}$ (\textbf{A}), Ca$^{2+}$ (\textbf{B}), Sr$^{2+}$ (\textbf{C}), F$^-$ (\textbf{D}), Cl$^-$ (\textbf{E}), and Br$^-$ (\textbf{F}). For Mg$^{2+}$, the 0.6~M condition shows inconsistent behavior compared to the other concentrations. The free energy decreases significantly below coordination number 6 due to a non-physical ion pairing event, likely as an artifact of the biased simulations. Here, we show the 0.55~M case, which is consistent with the other concentrations.}
    \label{fig:SI_all_fe_concentration}
\end{figure}

\clearpage
\pagebreak

\section{Correlations between coordination number, cross-sectional area, and shell volume}

 \begin{figure}[H]
    \centering
    \includegraphics[width=\textwidth]{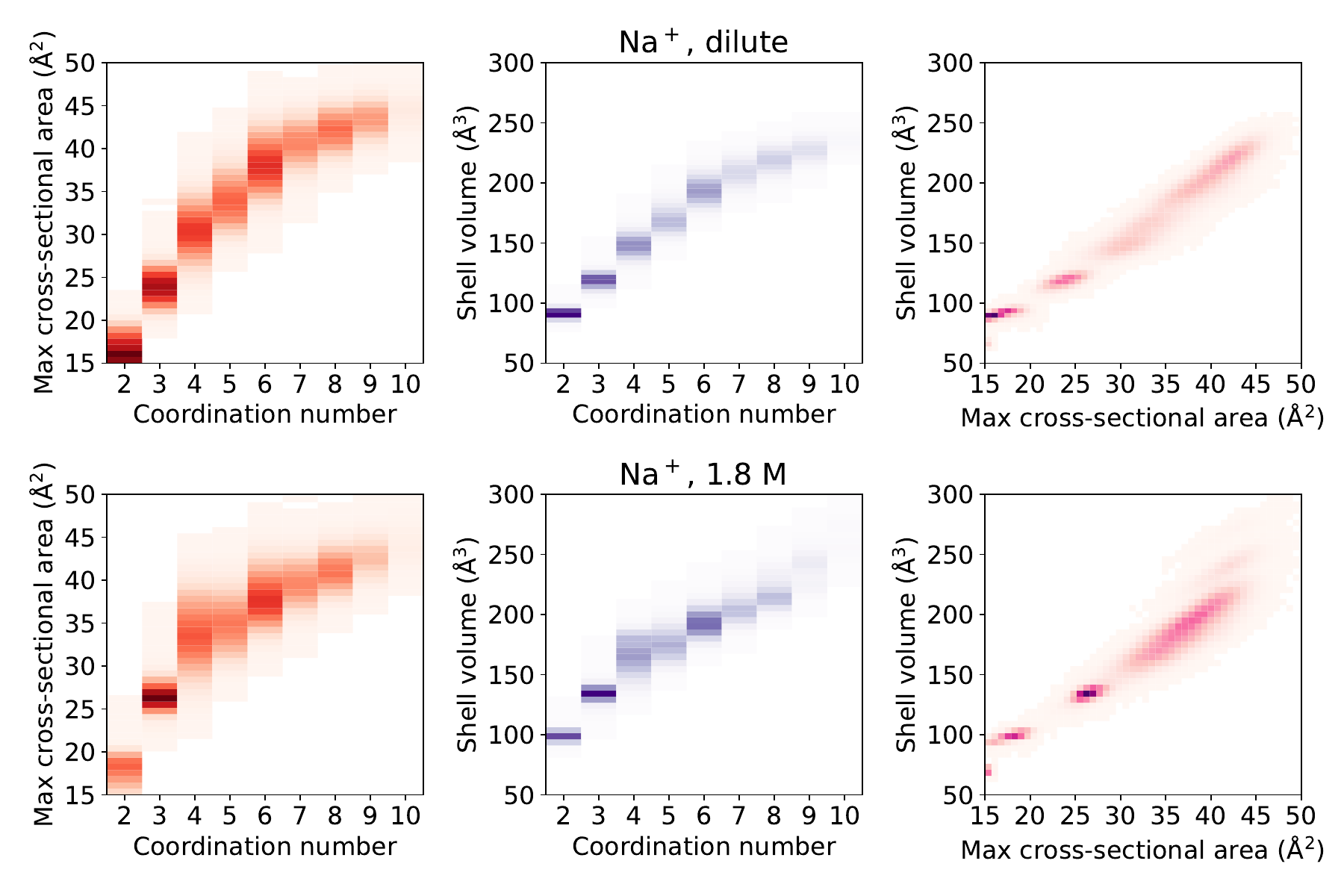}
    \caption{The restrictive cross-sectional area is more weakly correlated with the coordination number than the shell volume is. We show 2D histograms of the correlations among the coordination number, the limiting area and the shell volume for the umbrella sampling simulations of Na$^+$ at infinite dilution (top) and 1.8~M (bottom). At high salinity, there are more possible shell geometries for a given coordination number, since a given coordination number could be composed of a mix of ions and water molecules. The poor sampling in the cross-sectional area and shell volume between the low coordination number states may be an artifact of the biasing scheme. The biasing scheme requires a pre-defined hydration shell cutoff. However, these effects will not change the conclusions about geometric size constraints.}
    \label{fig:SI_correlations_area_volume_cn}
\end{figure}

\clearpage
\pagebreak

\section{Details on the calculation of cross-sectional area of the shell}\label{s:SI_limiting_area}

\begin{figure}[H]
    \centering
    \includegraphics[width=0.75\textwidth]{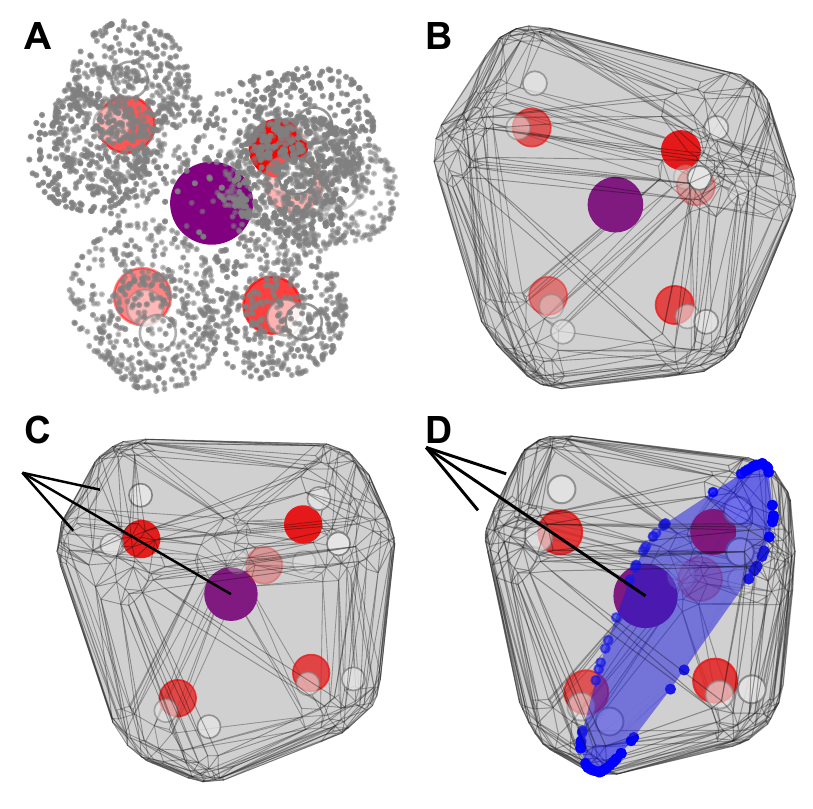}
    \caption{Visualization of the polyhedron formed by molecules in the shell. The convex hull formed by the van der Waals spheres of the shell is gray with edges darkened. The limiting cross-sectional area is blue, shown with the points intersecting the convex hull. The principal axis is shown as a black vector.}
    \label{fig:SI_polyhedron_visualization}
\end{figure}

\begin{itemize}[itemsep=1mm, parsep=0pt]
    \item Locate all atoms within hydration shell radius.
    \item Generate 3D point cloud on the surface of the van der Waals spheres for those molecules (Figure~\ref{fig:SI_polyhedron_visualization}A).
    \item Create polyhedron from point cloud using convex hull (Figure~\ref{fig:SI_polyhedron_visualization}B).
    \item Calculate the principal component of the vertices of the polyhedron (Figure~\ref{fig:SI_polyhedron_visualization}C).
    \item Create polygon from the plane intersecting the polyhedron, perpendicular to the principal component
    \item Repeat for the entire length of the principal axis
    \item Save the maximum cross-section along the principal axis (Figure~\ref{fig:SI_polyhedron_visualization}D).
\end{itemize}

\clearpage
\pagebreak

\section{Ion pairing distribution for Cl$^-$}

\begin{figure}[H]
    \centering
    \includegraphics[width=\textwidth]{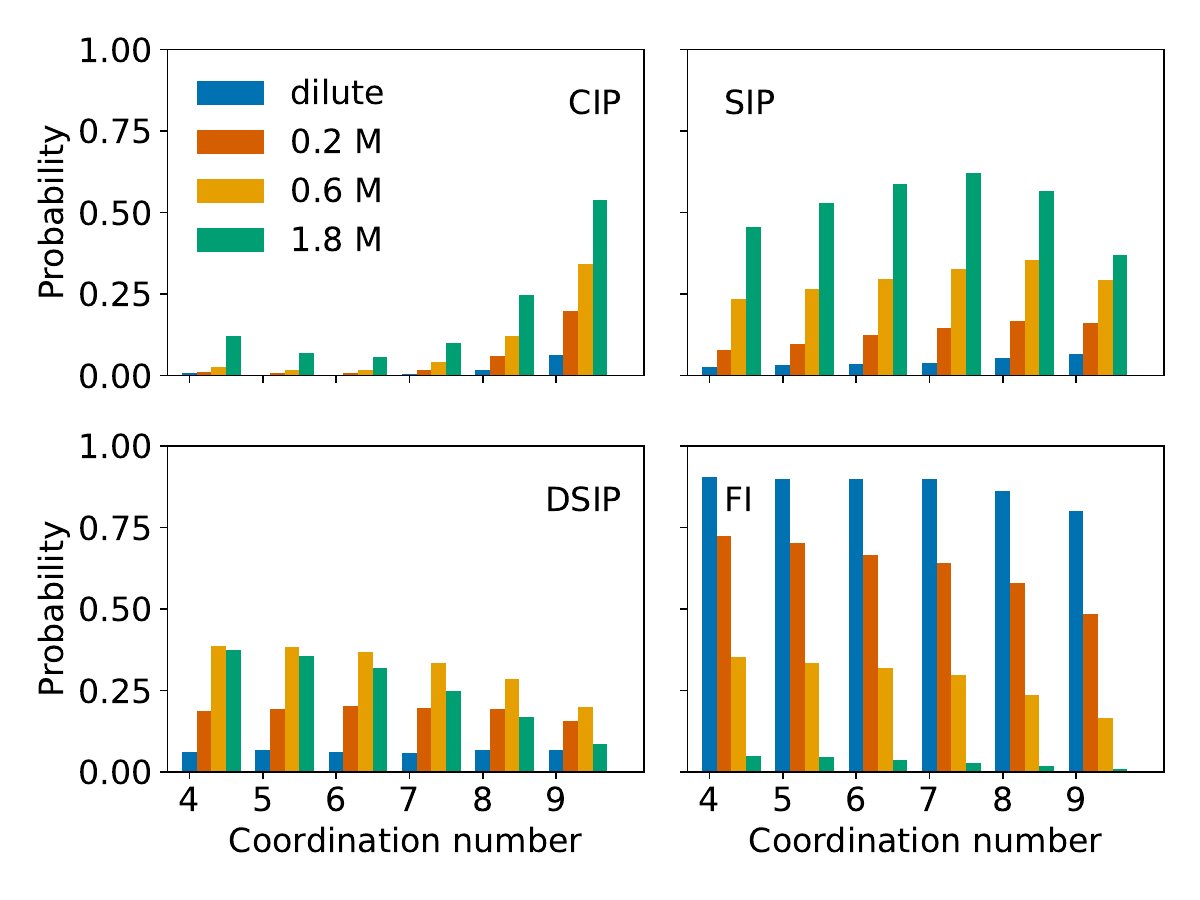}
    \caption{Cl$^-$ does not form ion pairs as frequently as 
    Na$^+$ upon de-coordination. Ion pairing probabilities for Cl$^-$ at 1~bar during the umbrella simulations. The frequencies are normalized by the number of frames in a given coordination number state, so the sum over the four ion pairing states for each coordination number is unity.}
    \label{fig:SI_Cl_ion_pairing}
\end{figure}

\clearpage
\pagebreak

\section{Umbrella simulation configurations} \label{s:SI_umbrella_configs}

Please refer to the attached machine-readable \texttt{tableS1.csv}.

\clearpage
\pagebreak


\end{document}